\documentclass[11pt,a4paper]{article}
\usepackage{jheppub}
\usepackage{amscd}
\usepackage{amsfonts}
\usepackage{amsthm}
\usepackage{stmaryrd}
\usepackage{tikz}
\usetikzlibrary{arrows,shapes}

\def\IC{\mathbb{C}}
\def\IN{\mathbb{N}}
\def\IZ{\mathbb{Z}}
\def\IR{\mathbb{R}}
\def\ID{\mathbb{D}}

\def\IJ{\mathbb{J}}
\def\IK{\mathbb{K}}

\def\IX{\mathbb{X}}

\def\IY{\mathbb{Y}}

\def\Ilb{\llbracket}
\def\Irb{\rrbracket}

\newcommand\cmp[3]{{\it Commun.\ Math.\ Phys.\ }{\bf #1} (#2) #3}
\newcommand\jhep[3]{{\it J. High Energy Phys.\ }{\bf #1} (#2) #3}

\newcommand\npb[3]{{\it Nucl.\ Phys.\ }{\bf B #1} (#2) #3}
\newcommand\plb[3]{{\it Phys.\ Lett.\ }{\bf B #1} (#2) #3}
\newcommand\mpla[3]{{\it Mod.\ Phys.\ Lett.\ }{\bf A #1} (#2) #3}
\newcommand\forp[3]{{\it Fortschr.\ Phys.\ }{\bf #1} (#2) #3}
\newcommand\hepth[1]{{\tt hep-th/}#1}

\newcommand{\pp}{{=\!\!\!|}}
\newcommand\fverb{\setbox\pippobox=\hbox\bgroup\verb}
\newcommand\fverbdo{\egroup\medskip\noindent%
            \fbox{\unhbox\pippobox}\ }
\newcommand\fverbit{\egroup\item[\fbox{\unhbox\pippobox}]}
\newbox\pippobox

\title{The generalized K{\"a}hler geometry of $N=(2,2)$ WZW-models}

\author[a]{Alexander Sevrin,}
\author[b]{Wieland Staessens}
\author[a]{and Dimitri Terryn}
\affiliation[a]{
Theoretische Natuurkunde, Vrije Universiteit Brussel\\ and\\The International Solvay Institutes\\
Pleinlaan 2, B-1050 Brussels, Belgium }
\affiliation[b]{Institut f\"ur Physik (WA THEP)\\
Johannes-Gutenberg-Universit\"at\\
Staudingerweg 7, D-55099 Mainz, Germany }

\emailAdd{Alexandre.Sevrin@vub.ac.be}
\emailAdd{Wieland.Staessens@uni-mainz.de}
\emailAdd{Dimitri.Terryn@vub.ac.be}

\abstract{$N=(2,2)$, $d=2$ supersymmetric non-linear $\sigma$-models provide a physical realization of Hitchin's and Gualtieri's generalized K\"ahler geometry. A large subclass of such models are comprised by WZW-models on even-dimensional reductive group manifolds. In the present paper we analyze the complex structures, type changing, the superfield content and the affine isometries compatible with the extra supersymmetry. The results are illustrated by an exhaustive discussion of the $N=(2,2)$ WZW-models on $S^3\times S^1$ and $S^3\times S^3$ where various aspects of generalized K\"ahler 
and Calabi-Yau geometry are verified and clarified. The examples illustrate a slightly weaker definition
for an $N=(2,2)$ superconformal generalized K\"ahler geometry compared to that for a generalized Calabi-Yau geometry.}


\begin{document}
\hspace{12cm}{MZ-TH/11-34}
 
\maketitle

\setcounter{equation}{0}

%
% Introduction
%

\section{Introduction} \label{introduction}
Hitchin's generalized K{\"a}hler geometry \cite{Hitchin:2004ut} is realized
by $N=(2,2)$ supersymmetric non-linear $\sigma $-models in two dimensions
\cite{Gualtieri:2003dx} (for reviews see \cite{reviews}). From the physics point of view, these models describe
strings propagating in a bihermitian background geometry
\cite{Gates:1984nk}, \cite{Howe:1985pm}. Indeed, requiring that an
$N=(1,1)$ supersymmetric non-linear $\sigma $-model has $N=(2,2)$
supersymmetry introduces
additional geometric structure, in particular two complex structures (one
for the left-handed and one for the right-handed extra supersymmetry transformations) such that the 
metric is hermitian with respect to both of them. The resulting conditions on the geometry can (locally)
be solved in terms of a single function \cite{Lindstrom:2005zr} which already
warrants the name {\em generalized K{\"a}hler geometry}. This finding was an
immediate consequence of the off-shell realization of the $\sigma $-model
in $N=(2,2)$ superspace \cite{Buscher:1987uw}-\cite{Bogaerts:1999jc}.

Imposing conformal invariance at the quantum level gives additional conditions which were
studied in the context of 1-loop $\beta$-functions in the nonlinear $\sigma$-model 
\cite{Grisaru:1997pg} and in 
the context of generalized Calabi-Yau geometry
\cite{Hull:2010sn}.

Till now there are not that many explicit -- in the sense that the
generalized K{\"a}hler potential is known -- non-trivial examples of
generalized K{\"a}hler geometries beyond usual K\"ahler geometry. An obvious example is provided by
hyper-K{\"a}hler manifolds which -- when both complex structures are chosen
to be equal -- can be formulated as an ordinary K{\"a}hler geometry. However
a hyper-K\"ahler manifold has an $S^2$ worth of complex structures.
Choosing for the left- and right-handed complex structures two
non-coinciding and non-antipodal points on the two-sphere gives a genuine non-trivial example
of generalized K\"ahler geometry. The generalized K{\"a}hler potential
encodes then not only the metric but the full two-sphere worth of complex
structures as well \cite{Sevrin:1996jr}. Another class of examples can be
obtained by deforming the above mentioned generalized K\"ahler structures
\cite{Gualtieri:2003dx}.

One of the most studied examples beyond (hyper-)K{\"a}hler geometry is the
Hopf surface $S^3\times S^1$ \cite{Rocek:1991vk}, \cite{Sevrin:1996jr},
\cite{Hull:2008vw}, \cite{Sevrin:2009na}. Even-dimensional
reductive Lie group manifolds -- of which the Hopf manifold is one as
$S^3\times S^1\simeq SU(2)\times U(1)$ -- provide a large class of
potential examples of this as the corresponding Wess-Zumino-Witten models
all allow for $N\geq(2,2)$ supersymmetry \cite{Spindel:1988nh}.

In the present paper we initiate a systematic study of the manifest
$N=(2,2)$ formulation of supersymmetric WZW-models. In the next section we introduce 
various properties of $N=(2,2)$ $d=2$ supersymmetry, its relation with generalized K\"ahler 
geometry and the manifest supersymmetric formulation in $N=(2,2)$ superspace. Section 3 
introduces $N=(2,2)$ supersymmetric WZW-models and discusses various aspects particular to 
this class of models. In section 4 we give an exhaustive description of the WZW-model on 
$SU(2)\times U(1)$ (or the Hopf surface $S^3\times S^1$). The full moduli space of complex 
structures is explored; isometries, global aspects and the relation between various formulations 
are discussed in detail. In section 5 we provide an additional explicit example illustrating further 
issues: the WZW-model on $SU(2)\times SU(2)$. 
In sections 4 and 5 we pay particular attention to type-changing, generalized
Calabi-Yau conditions and other items related to generalized K\"ahler geometry.
We end with conclusions and an outlook. Our 
conventions are summarized in the appendix.

\section{$N=(2,2)$ $\sigma $-models}
\subsection{From $N=(1,1)$ to $N=(2,2)$ supersymmetry}
An $N=(2,2)$ non-linear $\sigma $-model requires a bihermitian geometry
$\{{\cal M}, g, H, J_+,J_-\}$ defined by \cite{Gates:1984nk},
\cite{Howe:1985pm}:
\begin{itemize}
  \item An even dimensional (target) manifold ${\cal M}$ endowed with a metric $g$ and a closed 3-form $H$. Locally we introduce a 2-form
  $b$, $H=db$ with the gauge symmetry $b\simeq b+dk$ with $k$ a
  1-form.
  \item Two (integrable) complex structures $J_+$ and $J_-$  which are such that the metric is hermitian with
  respect to both of them:
\begin{eqnarray}
&& J_+^2=J_-^2=-{\bf 1}\,,\nonumber\\
&& [X,Y]+J_\pm[J_\pm X,Y]+J_\pm[X,J_\pm Y]-[J_\pm X,J_\pm Y]=0\,,\nonumber\\
&& g\big(J_\pm X, J_\pm Y\big)=g(X,Y)\,,
\end{eqnarray}
where $X$ and $Y$ are vectors.
\item The hermiticity of the metric implies
the existence of two 2-forms $\omega _+$ and $\omega _-$,
\begin{eqnarray}
 \omega _\pm=-g(X,J_\pm Y).
\end{eqnarray}
Their exterior derivative should be given by,
\begin{eqnarray}
 d\omega _\pm\big(X,Y,Z\big)=\mp H(J_\pm X,J_\pm Y,J_\pm Z),\label{cddd}
\end{eqnarray}
where $X$, $Y$ and $Z$ are vectors.
\end{itemize}
These conditions have various consequences. One of those is that the
  complex structures are covariantly constant though with different
  connections $\Gamma_\pm$ (known as the Bismut connections),
\begin{eqnarray}
\Gamma_\pm \equiv  \left\{ \right\} \mp \frac 1 2
H\,,\label{cons}
\end{eqnarray}
where $\left\{ \right\}$ is the standard Levi-Civita connection. In fact taking the other conditions
into account, eq.~(\ref{cddd}) and $\nabla^{(\pm)}J_\pm=0$ are easily shown to be equivalent.

\subsection{Generalized K\"ahler geometry}
The geometric structure defined above is also known as a generalized
K{\"a}hler structure\footnote{Most of our discussion will be restricted to local issues only. A proper
global treatment requires the introduction of gerbes. For more details see {\em e.g.} the last reference in
\cite{reviews}.}. Indeed, given a manifold $\cal M$, Hitchin considered the bundle
$T\oplus T^\ast$, where $T$ and $T^\ast$ are the tangent and the cotangent
bundle resp., on which one defines a natural symmetric bilinear pairing,
\begin{eqnarray}
\langle \IX, \IY\rangle  = \frac 1 2 \left( \iota_X\eta + \iota_Y\xi \right) ,\label{np}
\end{eqnarray}
where $\IX=X+\xi,\,\IY=Y+\eta\,\in T\oplus T^\ast$.
In Hitchin's geometry, the role of the Lie bracket is replaced by the
$H$-twisted Courant bracket,
\begin{eqnarray}
\Ilb \IX,\IY\Irb_H = [X,Y] + {\cal L}_X \eta - {\cal L}_Y \xi - \frac 1 2 d \big(\iota_X\eta- \iota_Y\xi\big)
+\iota_Y \iota_XH,\label{Courant}
\end{eqnarray}
where $H\in\wedge^3T^\ast$ is a closed 3-form\footnote{We define the interior product so that
the vector is always contracted with the first argument of the form. {\em I.e.} $\iota_X \omega= 
\omega(X,\cdots)\,$.}. While anti-symmetric, the Courant bracket does not satisfy the Jacobi identities,
\begin{eqnarray}
&&\Ilb\IX,\Ilb \IY,\IZ\Irb\Irb_H+\Ilb\IY,\Ilb \IZ,\IX\Irb\Irb_H+
\Ilb\IZ,\Ilb \IX,\IY\Irb\Irb_H=
-\frac 1 3\,d \big(
\langle \IX, \Ilb \IY,\IZ\Irb_H\rangle +\langle \IY, \Ilb \IZ,\IX\Irb_H\rangle +
\nonumber\\
&&\qquad\qquad\qquad\langle \IZ, \Ilb \IX,\IY\Irb_H\rangle\big)\,,
\label{jacobi}
\end{eqnarray}
where $\IX=X+\xi,\,\IY=Y+\eta,\,\IZ=Z+\zeta\in T+T^\ast$ and where we used $dH=0$. However it is clear that the Jacobi identities are satisfied on an isotropic subspace\footnote{An isotropic subspace $L \subset T\oplus T^\ast$ is defined by
$\forall \,\IX,\,\IY\in L: \langle\IX,\IY\rangle =0$. If the dimension of $T\oplus T^\ast$ is $2m$, then the maximal dimension of $L$ is given by $m$. Whenever the dimension of $L$ is $m$, we talk about a maximal isotropic subspace.} of $T\oplus T^\ast$ provided the Courant bracket acts involutively on the isotropic subspace.

The bilinear form eq.~(\ref{np})
has a large isometry group of which the so-called $b$-transformations form an
important subgroup,
\begin{eqnarray}
\IX \rightarrow  e^b \,\IX=\IX+\iota_X b\,,\label{btransfX}
\end{eqnarray}
where $b$ is a locally defined two-form. It
is then straightforward to show that,
\begin{eqnarray}
\Ilb e^b (\IX), e^b (\IY)\Irb_H = e^b \Ilb\IX,\IY\Irb_{H +db}\,.
\end{eqnarray}
So the Courant bracket is invariant provided $b$ is closed. In fact, Hitchin's geometry
extends the usual Lie derivative action on $T\oplus T^\ast$ such that the $b$-transformation 
is included as well. One gets,
\begin{eqnarray}
\hat{\cal L}_\IX\,\IY\equiv [X,Y]+ {\cal L}_X\eta+\iota_Y(\iota_XH-d \xi)\,,\label{genLie}
\end{eqnarray}
where the first two terms correspond to the usual Lie derivatives and the last term is 
a $b$-transformation with $b=\iota_XH-d\xi$. With this one verifies,
\begin{eqnarray}
\hat{\cal L}_\IX\,\IY-\hat{\cal L}_\IY\,\IX=2\,\Ilb \IX,\IY\Irb {}_H\,.
\end{eqnarray}

The exterior algebra on $T^\ast$ provides a natural choice for spinors in a generalized geometry. 
For each $\IX=X+\xi\in T\oplus T^\ast$ we introduce $ \Gamma_\IX$ which acts on
$\phi\in \wedge^\bullet T^\ast$ as,
\begin{eqnarray}
\Gamma_\IX\cdot\phi=\iota_X\phi+\xi\wedge \phi\,,
\end{eqnarray}
and with this one verifies that,
\begin{eqnarray}
\big\{ \Gamma_\IX, \Gamma_\IY\big\}\cdot\phi=2\langle \IX,\IY\rangle\,\phi\,.\label{dirac}
\end{eqnarray}
In the standard way this yields the spin representation of the isometry group of the bilinear form. 
In particular we get 
that the $b$ transform, see eq.~(\ref{btransfX}),  acts as,
\begin{eqnarray}
\phi \rightarrow e^{-b\wedge}\,\phi\,.\label{bspin}
\end{eqnarray}
In a similar way one finds how pure spinors transform under coordinate transformations. They do not 
transform as an element of $ \wedge^\bullet T^\ast$ but rather as a density:
\begin{eqnarray}
x \rightarrow x'(x)\Rightarrow \phi(x) \rightarrow \phi'(x')=\sqrt{\det \frac{\partial x'}{\partial x}}\, 
\phi(x)\,.
\end{eqnarray}
The Lie derivative action eq.~(\ref{genLie}) generalizes to spinors,
\begin{eqnarray}
\hat{\cal L}_\IX\, \phi&\equiv&\big(d-H\wedge\big) \Gamma_\IX\cdot \phi+
\Gamma_\IX\cdot \big(d-H\wedge\big) \phi \nonumber\\
&=& {\cal L}_X \phi-(\iota_XH-d \xi)\wedge \phi\,,
\end{eqnarray} 
which using eq.~(\ref{bspin}) is clearly compatible with eq.~(\ref{genLie}).  
On the spinors the Mukai pairing gives an invariant (under the isometry group of $\langle\cdot,
\cdot\rangle$ connected to the identity) bilinear form,
\begin{eqnarray}
\big( \phi_1,\phi_2\big)= \sigma( \phi_1)\wedge \phi_2|_{\mbox{top}}\,,
\end{eqnarray}
where $\phi_1,\, \phi_2\in\wedge^\bullet T^\ast$ and $\sigma$ reverses the order of the basis of 
the forms, {\em i.e.},
\begin{eqnarray}
\sigma\big(dx^1\wedge dx^2\wedge \cdots\wedge dx^r\big)=
dx^r\wedge dx^{r-1}\wedge \cdots\wedge dx^1\,.
\end{eqnarray}
There is a natural way to associate an isotropic subspace $L$ of $T\oplus T^\ast$ to a given spinor 
$\phi$:
\begin{eqnarray}
\IX\in L \Leftrightarrow \Gamma_\IX\cdot\phi=0\,.
\end{eqnarray}
Using eq.~(\ref{dirac}) one immediately shows that $L$ is indeed isotropic. When $L$ is maximally 
isotropic one calls $\phi$ a {\em pure spinor}.

There are several equivalent definitions for a {\em generalized complex structure}. 
We choose the one that 
is quite close to the usual definition of a complex structure. An $H$-twisted generalized 
complex structure (HGCS) is defined as a
linear map ${\cal J}:  T \oplus T^\ast \rightarrow T \oplus T^\ast$,
satisfying ${\cal J}^2 = -1$, which preserves the natural pairing,
$\langle {\cal J} \IX, {\cal J} \IY \rangle = \langle \IX, \IY \rangle$
for all $\IX, \IY \in T \oplus T^\ast$ and for which the $+i$ eigenbundle is involutive under the 
H-twisted Courant bracket. The latter can be seen as an integrability condition 
similar to the the requirement that the Nijenhuis tensor vanishes, but
now the Lie bracket is replaced by the $H$-twisted Courant bracket,
\begin{eqnarray}
 \Ilb \IX,\IY \Irb_H +{\cal J}\Ilb {\cal J} \IX,\IY\Irb_H +{\cal J}\Ilb \IX,{\cal J} \IY\Irb_H -\Ilb{\cal J} 
\IX,{\cal J} \IY\Irb_H =0.
\end{eqnarray}
A necessary requirement for a HGCS to exist is that the manifold $\cal M$ is even dimensional,
so from now on we take $\dim {\cal M}=2m$.
Note that if ${\cal J}$ is $H$-twisted, then $e^{-b} {\cal J} e^{b}$ is
$(H+db)$-twisted. Writing $(T\oplus T^\ast)\otimes \IC=L\oplus \bar L$ where $L$ ($\bar L$) is the
$+i$ ($-i$) eigenbundle of $ {\cal J}$, we see that $L$ is a maximal isotropic subspace
of $(T\oplus T^\ast)\otimes \IC$. As a consequence, any HGCS comes with a pure spinor $\phi$ defined
by $ \Gamma_\IX\cdot\phi=0$, $\forall\, \IX\in L$. In a physicists language this is just the highest weight
vector of the spinor representation. 

Canonical examples of a GCS (with $H=0$) are complex structures and symplectic structures. Take 
for instance  a 
complex structure $J$ on $T$ and $\omega\in\wedge^2T^\ast$ a closed non-degenerate two-form,
then both,
\begin{eqnarray}
{\cal J}_c =  \left( \begin{array}{cc}
                                        J & 0 \\
                                        0 & -J^t
                                        \end{array}\right)\,,
\qquad                                        
{\cal J}_s = \left( \begin{array}{cc}
                                        0 & \omega^{-1} \\
                                        -\omega & 0
                                        \end{array}\right)\,,\label{canpusp}
\end{eqnarray}
are generalized complex structures. For ${\cal J}_c$ the $+i$ eigenspace consists of 
$\IX\in T^{(1,0)}\oplus T^\ast_{(0,1)}$ and the associated pure spinor is $\phi\in\wedge^{(0,m)} 
T^\ast$. For ${\cal J}_s$ the $+i$ eigenspace is given by $\IX=X-i\,\iota_X\omega$ with the 
corresponding pure spinor $\phi=e^{i\, \omega\wedge}\,$.

A generic $HGCS$ interpolates between the two extreme cases, complex and symplectic structures. 
The Newlander-Nirenberg and the Darboux theorems respectively guarantee the existence of natural local 
coordinates for a complex and a symplectic structure respectively. Gualtieri extended this to an arbitrary
HGCS \cite{Gualtieri:2003dx}. By an appropriate diffeomorphism and $b$-transformation one can 
always turn a HGCS to the standard product GCS $\IC^k\times (\IR^{2m-2k}, \omega)$. The integer 
$k$ is called the {\em type} of the HGCS. 
In this sense, a $2m$ dimensional manifold with a generalized 
complex structure is foliated by $2m-2k$ dimensional leaves of the form 
$\IR^{2m-2k} \times  \{ {\rm 
point}\}$ on which a symplectic form $\omega$ can be properly defined. Transverse to the leaves, 
we can introduce complex coordinates $z_i$ with $i \in \{ 1, \ldots , k\}$, such that the leaves are 
located at $z_i = {\rm constant} \, (\forall \, i)$.  Gualtieri's theorem only holds for neighborhoods of 
regular points. A generic feature of generalized complex geometry is that loci might exist where the 
type jumps, one calls this phenomenon {\em type changing}.

A {\em generalized $H$-twisted K{\"a}hler structure} is a set of two mutually
commuting HGCS's, ${\cal J}_+$ and ${\cal J_-}$ such that,
\begin{eqnarray}
{\cal G}\big(\IX,\IY\big)=\langle {\cal J}_+\IX, {\cal J}_- \IY\rangle,
\end{eqnarray}
defines a positive definite metric on $T\oplus
T^*$. Gualtieri \cite{Gualtieri:2003dx} showed that for $H=db$ (locally), a
generalized $H$-twisted K{\"a}hler structure is precisely equivalent to the bihermitian
geometry which follows from a $N=(2,2)$ non-linear $\sigma $-model (as was introduced in section 2.1),
where,
\begin{eqnarray}
{\cal J}_{\pm} = \frac 1 2\,       \left( \begin{array}{cc}
                                        J_+ \pm J_- & \omega^{-1}_+ \mp \omega^{-1}_- \\
                                        -(\omega_+ \mp \omega_-) & -(J^t_+ \pm J^t_-)
                                        \end{array}\right)\,,
                                        \label{gcs221}
\end{eqnarray}
and where the Courant bracket is $H$-twisted. Untwisting the Courant bracket ({\em i.e.} taking $H=0$ in 
eq.~(\ref{Courant})) one gets,
\begin{eqnarray}
{\cal J}_{\pm} = \frac 1 2\left( \begin{array}{cc}1&0\\-b&1 \end{array}\right)\,
                                        \left( \begin{array}{cc}
                                        J_+ \pm J_- & \omega^{-1}_+ \mp \omega^{-1}_- \\
                                        -(\omega_+ \mp \omega_-) & -(J^t_+ \pm J^t_-)
                                        \end{array}\right)\,
                                        \left( \begin{array}{cc}1&0\\+b&1 \end{array}\right)\,.
                                        \label{gcs22}
\end{eqnarray}
This is the Gualtieri map.

The simplest example is ordinary K\"ahler geometry where we choose $J_+=J_-=J$, $b=0$ and we get
that ${\cal J}_+={\cal J}_c$ and ${\cal J}_-={\cal J}_s$ where $\omega$ is the K\"ahler form,
$\omega(X,Y)=-g(X,JY)$. Another interesting example, first
noted in \cite{Sevrin:1996jr} and worked out in \cite{Gualtieri:2003dx}, is 
provided by hyper-K\"ahler manifolds. We denote the complex structures by $J_a$, 
$a\in\{1,2,3\}$, satisfying $J_aJ_b=-\delta_{ab}+ \varepsilon_{abc}J_c$. We call the
three K\"ahler two-forms $\omega_a$, $\omega_a(X,Y)=-g(X,J_aY)$. While this is a 
K\"ahler manifold, one can also choose\footnote{A hyper-K\"ahler manifold has an 
$S^2$ of complex structures. Identifying $J_+$ and $J_-$ with two non-coinciding, non antipodal 
points leads to similar results.} $J_+=J_1$ and $J_-=J_2$ leading to the situation where $\ker[J_+,J_-]=0$. The generalized K\"ahler structure is then given by,
\begin{eqnarray}
{\cal J}_{\pm} &=& \left( \begin{array}{cc}1&0\\ \mp \omega_3&1 \end{array}\right)\,
                                        \left( \begin{array}{cc}
                                        0& \frac 1 2 \big(\omega^{-1}_1 \mp \omega^{-1}_2\big) \\
                                        -(\omega_1 \mp \omega_2) & 0
                                        \end{array}\right)\,
                                        \left( \begin{array}{cc}1&0\\\pm \omega_3&1 \end{array}\right),
\label{gualt}
\end{eqnarray}
{\em i.e.} both ${\cal J}_+$ and $ {\cal J}_-$ are of symplectic type.

For generalized K\"ahler geometries we have two generalized complex structures ${\cal J}_+$
and
${\cal J}_-$ and 
with each generalized complex structure we can associate a type, $k_+$ and $k_-$ resp. From 
eq.~(\ref{gcs22})   one finds that   
%:tt
\begin{eqnarray}
 k_\pm \equiv{\rm type} \, ({\cal J}_\pm) = \frac{1}{2} {\rm corank}_{\IR}(\omega_+^{-1}\mp 
\omega_-^{-1})= \frac{1}{2} \big( 2m -  {\rm rank}_{ \IR} (\omega_+^{-1}\mp \omega_-^{-1})  
\big)\,.
\end{eqnarray}    
We call $(k_+,k_-)$ the type of the generalized K\"ahler geometry. So we can also write,
\begin{eqnarray}
\big(k_+,k_-\big)=\frac 1 2 \,\big(\dim\ker(J_+-J_-),\dim\ker(J_++J_-)\big).
\end{eqnarray}

As $ {\cal J_+}$ and $ {\cal J_-}$ commute, we can write,
\begin{eqnarray}
\big(T\oplus T^\ast\big)\otimes \IC=
L_{++}\oplus L_{+-}\oplus L_{--}\oplus L_{-+},
\end{eqnarray}
where $L_{++}$ is the $+i$ eigenbundle for both $ {\cal J_+}$ and $ {\cal J_-}$; 
$L_{+-}$ is the $+i$ eigenbundle for $ {\cal J_+}$ and the $-i$ eigenbundle for $ {\cal J_-}$; etc.
Using the Gualtieri map eq.~(\ref{gualt}) and the hermiticity of the metric w.r.t. both $J_+$ and $J_-$
one verifies,
\begin{eqnarray}
&&\IX_+\in L_{++} \Leftrightarrow \IX_+=X_++(g-b)X_+{\mbox{ and }}
\frac 1 2 (1-iJ_+)X_+=X_+\,, \nonumber\\
&&\IX_-\in L_{+-} \Leftrightarrow \IX_-=X_--(g+b)X_-{\mbox{ and }}
\frac 1 2 (1-iJ_-)X_-=X_-\,, \nonumber\\
&&\bar\IX_+\in L_{--} \Leftrightarrow \bar\IX_+=\bar X_++(g-b)\bar X_+{\mbox{ and }}
\frac 1 2 (1+iJ_+)\bar X_+=\bar X_+\,, \nonumber\\
&&\bar\IX_-\in L_{-+} \Leftrightarrow \bar\IX_-=\bar X_--(g+b)\bar X_-{\mbox{ and }}
\frac 1 2 (1+iJ_-)\bar X_-=\bar X_-\,.\label{eigenbundle}
\end{eqnarray}
We can now introduce pure spinors $\phi_+$ and $\phi_-$ for $ {\cal J_+}$ and $ {\cal J_-}$ resp. 
They are defined by,
\begin{eqnarray}
\Gamma_{\IX_+}\cdot\phi_+=\Gamma_{\IX_-}\cdot\phi_+=0,\qquad
\Gamma_{\IX_+}\cdot\phi_-=\Gamma_{\bar\IX_-}\cdot\phi_-=0.\label{psdef}
\end{eqnarray}
Eq.~(\ref{psdef}) does not fix the normalization of the pure spinors $\phi_+$ and $\phi_-$. 
In \cite{Gualtieri:2003dx} it was shown that the integrability of the generalized complex structures 
guarantees the existence of $\IY_+$ and $\IY_-$ such that\footnote{If instead of eq.~(\ref{gcs22})  we 
would have taken the $H$-twisted generalized complex structures in eq.~(\ref{gcs221}), 
this relation would change to 
$(d-H\wedge) \phi_\pm= \Gamma_{\IY_\pm}\cdot \phi_\pm$, {\em i.e.} the exterior derivative is replaced by an $H$-twisted analogue which is still nilpotent as $H$ is a closed 3-form.}
\begin{eqnarray}
d \phi_\pm= \Gamma_{\IY_\pm}\cdot \phi_\pm\,.
\end{eqnarray}
Explicit solutions for the pure spinors were provided in \cite{Halmagyi:2007ft} (for the case $d=6$ where
$\mbox{im}\big( [J_+,J_-]g^{-1}\big)\neq 0$) 
and in \cite{Hull:2010sn} (for 
the general case). We will come back to this in the next section.

%:dfdf

Finally, a {\em generalized Calabi-Yau geometry} is a generalized K\"ahler geometry for which the
pure spinors $\phi_+$ and $\phi_-$ are {\em globally defined}, {\em closed} and they satisfy,
\begin{eqnarray}
\big(\phi_+,\bar \phi_+\big)= c\, \big(\phi_-,\bar \phi_-\big)\neq 0,\label{gcycd}
\end{eqnarray}
where we denoted the complex conjugate of $\phi$ by $\bar \phi$ and where $c$ is some non-zero 
constant. In the current definition we consider a generalized K\"ahler geometry where the complex
structures are integrable with respect to the untwisted Courant bracket. If instead we would consider 
a generalized K\"ahler geometry with respect to the twisted Courant bracket, the definition of a generalized Calabi-Yau geometry would remain unchanged except that the pure spinors are now not closed anymore but $H$-closed,
\begin{eqnarray}
d\, \phi_\pm=H\wedge \phi_\pm.
\end{eqnarray}
There is also some ambiguity in the current definition of a generalized Calabi-Yau geometry. Indeed we 
saw before that under a coordinate transformation a pure spinor does not transform as an element of  
$\wedge^\bullet T^\ast$ but as a density. This is {\em e.g.} necessary to keep the Mukai pairing invariant
under coordinate transformations. This however makes the notion of a closed pure spinor a coordinate
dependent statement. Once again we will come back to this in the next sections.

\subsection{Supersymmetric $\sigma$-models in $N=(2,2)$ superspace}
As a full off-shell description of $N=(2,2)$ non-linear $\sigma $-models
in $N=(2,2)$ superspace is known \cite{Lindstrom:2005zr} (see also
\cite{Sevrin:1996jr} and \cite{Bogaerts:1999jc}), this geometry can be
locally characterized in terms of a single real potential (the Lagrange
density)\footnote{Introducing boundaries for the non-linear
$\sigma$-model -- relevant for the description of open strings in an
NSNS-background -- reduces the $N=(2,2)$ supersymmetry to an $N=2$
supersymmetry. Besides the bulk geometry which is locally determined by
the generalized K{\"a}hler potential, one now gets a second real potential
which lives on the boundary and which is (partially) determined by the
boundary conditions \cite{Sevrin:2009na}.}. The construction starts from
the observation that the off-shell non-closing terms in the $N=(2,2)$
supersymmetry algebra in $N=(1,1)$ superspace are all proportional to the
commutator of the two complex structures $[J_+,J_-]$. Therefore in order
to get off-shell closure one expects additional auxiliary fields in
the direction of $\mbox{im}\big({[}J_+,J_-{]}g^{-1}\big)$ while this will not be
needed for $\ker[J_+,J_-]=\ker(J_+-J_-)\oplus\ker(J_++J_-)$.

Decomposing the tangent space as
$T=\ker(J_+-J_-)\oplus\ker(J_++J_-)\oplus\mbox{im}\big({[}J_+,J_-{]}g^{-1}\big)$ one
indeed gets (conjectured in \cite{Sevrin:1996jr} and proven in
\cite{Lindstrom:2005zr}) that the first subspace gets parameterized by
chiral, the second by twisted chiral \cite{Gates:1984nk} and the last one
by semi-chiral $N=(2,2)$ superfields \cite{Buscher:1987uw}. The three
types of superfields are defined by the following constraints\footnote{We
refer to the appendix for our conventions. We make a distinction between
letters from the beginning ($\alpha ,\,\beta ,\,\gamma,\,$...) and
letters from the middle of the Greek alphabet ($\mu ,\,\nu ,\,\rho
,\,$...)}:
\begin{description}
  \item[Semi-chiral superfields:] $l^{\tilde \alpha }$, $l^{\bar{ \tilde
  \alpha}  }$, $r^{\tilde \mu  }$, $r^{\bar{ \tilde \mu }  }$,
  $\qquad\tilde \alpha,\,\bar{ \tilde \alpha} ,\,\tilde \mu ,\,\bar{
  \tilde \mu }\in \{1,\cdots n_s\}$,
\begin{eqnarray}
\bar \ID_+ l^{\tilde \alpha }=\ID_+l^{\bar{ \tilde
  \alpha}  }= \bar \ID_-r^{\tilde \mu  }=\ID_-r^{\bar{ \tilde \mu }  }=0.
\end{eqnarray}
  \item[Twisted chiral superfields:] $w^\mu $, $w^{\bar \mu }$, $\qquad\mu,\,\bar \mu
  \in\{1,\cdots n_t\}$,
\begin{eqnarray}
 \bar \ID_+w^\mu=\ID_-w^\mu=\ID_+w^{\bar \mu }=\bar \ID_-w^{\bar \mu }=0.
\end{eqnarray}
  \item[Chiral superfields:] $z^\alpha $, $z^{\bar \alpha }$, $\qquad\alpha,\, \bar \alpha
  \in\{1,\cdots n_c\}$,
\begin{eqnarray}
 \bar \ID_\pm z^\alpha =\ID_\pm z^{\bar \alpha }=0.
\end{eqnarray}
\end{description}
Chiral and twisted chiral $N=(2,2)$ superfields have the same number of
component fields as $N=(1,1)$ superfields while semi-chiral $N=(2,2)$
superfields have twice as many, half of which are -- from $N=(1,1)$
superspace point of view -- auxiliary.

The most general action involving these superfields and consistent with
dimensions is given by,
\begin{eqnarray}
{\cal S}=4\,\int\,d^2 \sigma \,d^2\theta \,d^2 \hat \theta \, V(l,\bar l,r,\bar r,w,\bar w, z, \bar z),
\label{actionN22}
\end{eqnarray}
where the Lagrange density $V(l,\bar l,r,\bar r,w,\bar w, z, \bar z)$ is
an arbitrary real function of the semi-chiral, the twisted chiral and the
chiral superfields. It is defined modulo a generalized K{\"a}hler
transformation,
\begin{eqnarray}
V\rightarrow V+F(l,w,z)+ \bar F(\bar l,\bar w,\bar z)+ G(\bar r,w,\bar z)+ \bar G(r,\bar w,z).\label{genKahltrsf1}
\end{eqnarray}
As for the usual K{\"a}hler case these generalized K{\"a}hler transformations
are essential for the global consistency of the model, see {\em e.g.} the
seminal work in \cite{Hull:2008vw}. Before proceeding, we introduce some
notation. We write,
\begin{eqnarray}
 M_{AB}=\left( \begin{array}{cc}
   V_{ab} & V_{a\bar b} \\
   V_{\bar a b} & V_{\bar a\bar b}
 \end{array}\right),
\end{eqnarray}
where, $(A,a)\in\{(l,\tilde \alpha ),\,(r,\tilde \mu ),\,(w,\mu
),\,(z,\alpha )\}$ and $(B,b)\in\{(l,\tilde \beta ),\,(r,\tilde \nu
),\,(w,\nu ),\,(z,\beta )\}$. The subindices on $V$ denote derivatives
with respect to those coordinates. In this way {\em e.g.} we get that
$M_{zr}$ is the $2n_c\times 2n_s$ matrix given by,
\begin{eqnarray}
 M_{zr}=\left( \begin{array}{cc}
   V_{\alpha \tilde \nu } & V_{\alpha \bar{\tilde \nu }} \\
   V_{\bar \alpha \tilde \nu } & V_{\bar \alpha \bar{\tilde \nu }}
 \end{array}\right).
\end{eqnarray}
Note that $M_{AB}^T=M_{BA}$. We also introduce the matrix $\IJ$,
\begin{eqnarray}
 \IJ\equiv i\,\left(
\begin{array}{cc}
  {\bf 1} & 0 \\
  0 & -{\bf 1}
\end{array}
 \right),
\end{eqnarray}
with $\bf 1$ the unit matrix and using this we write,
\begin{eqnarray}
 C_{AB}\equiv \IJ\, M_{AB}-M_{AB}\,\IJ,\qquad
A_{AB}\equiv \IJ\, M_{AB}+M_{AB}\,\IJ. \label{canda}
\end{eqnarray}

Upon passing to $N=(1,1)$ superspace and after elimination of the
auxiliary fields -- one gets the complex structures,
\begin{eqnarray}
 J_+&=&\left(\begin{array}{cccc}
   \IJ & 0 & 0 & 0 \\
   M_{lr}^{-1}C_{ll} &  M_{lr}^{-1}\IJ M_{lr} &
   M_{lr}^{-1}C_{lw}&M_{lr}^{-1} C_{lz}   \\
   0 & 0 & \IJ & 0 \\
   0 & 0 & 0 & \IJ
 \end{array}   \right),\nonumber\\
J_-&=&\left(\begin{array}{cccc}
   M_{rl}^{-1}\IJ M_{rl} &  M_{rl}^{-1} C_{rr} &
   M_{rl}^{-1}A_{rw}&M_{rl}^{-1} C_{rz}    \\
0&\IJ  & 0 & 0 \\
   0 & 0 & -\IJ & 0 \\
   0 & 0 & 0 & \IJ
 \end{array}   \right),\nonumber\\
\end{eqnarray}
where we labeled rows and columns in the order $l$, $\bar l$, $r$, $\bar
r$, $w$, $\bar w$, $z$, $\bar z$. Note that once semi-chiral fields are
present, neither of the complex structures is diagonal. One easily shows
\cite{Sevrin:1996jr} that making a coordinate transformation which
replaces $r^{\tilde \mu }$ and $r^{\bar {\tilde \mu} }$ by $V_{\tilde
\alpha }$ and $V_{\bar{\tilde \alpha} }$ resp.~while keeping the other
coordinates as they are, diagonalizes $J_+$. Similarly, a coordinate
transformation which goes from $l^{\tilde \alpha }$ and $l^{\bar {\tilde
\alpha} }$ to $V_{\tilde \mu }$ and $V_{\bar{\tilde \mu} }$ and keeping
the other coordinates fixed diagonalizes $J_-$. This led to a
reinterpretation of the generalized K{\"a}hler potential as a generating
function for canonical transformations which interpolate between the
coordinate system in which $J_+$ is diagonal and the one in which $J_-$
is diagonal \cite{Lindstrom:2005zr}.

{}From the second order action one reads off the metric $g$ and the
torsion two-form potential $b$. One finds two natural expressions,
\begin{eqnarray}
&&\big(g-b_+\big)(X,Y)=\Omega ^+(X,J_+Y)=d B^+\,(X,J_+Y)\,, \nonumber\\
&&\big(g+b_-\big)(X,Y)=\Omega ^-(X,J_-Y)=dB^-\,(X,J_-Y)\,.\label{gminb}
\end{eqnarray}
where $\Omega^+$ and $\Omega^-$ are two (locally defined) closed two-forms linear in the
generalized K{\"a}hler potential and $H=db_+=db_-$. From eq.~(\ref{gminb}) one sees that $b_+$
and $b_-$ resp.~are $(2,0)$+$(0,2)$ forms w.r.t. $J_+$ and $J_-$ resp.  
Explicitly we get,
\begin{eqnarray}
\Omega^+ = -\frac 1 2 \left(\begin{array}{cccc} C_{ll} & A_{lr} &
C_{lw} & A_{lz}\\
-A_{rl} & -C_{rr} & -A_{rw} & -C_{rz} \\
C_{wl}& A_{wr} & C_{ww} & A_{wz} \\
-A_{zl} & -C_{zr} & -A_{zw} & -C_{zz}
\end{array}
 \right), \label{omplus}
 \end{eqnarray}
\begin{eqnarray}
\Omega^- = \frac 1 2 \left(\begin{array}{cccc} C_{ll} & C_{lr} &
C_{lw} & C_{lz}\\
C_{rl} & C_{rr} & C_{rw} & C_{rz} \\
C_{wl}& C_{wr} & C_{ww} & C_{wz} \\
C_{zl} & C_{zr} & C_{zw} & C_{zz}
\end{array}
 \right).\label{ommin}
\end{eqnarray}
where we labeled rows and columns in the order $(l, \bar l, r, \bar r,
w, \bar w,z,\bar z)$. Locally we can write $\Omega^\pm=dB^\pm$ where\footnote{The notation
we use is such that $V_l\,dl$ stands for $\partial_{l^{\tilde{ \alpha}}}  V\,dl^{\tilde{ \alpha}}\,$, $V_{\bar l}\,d\bar l$ for $\partial_{l^{\bar{\tilde{ \alpha}}}}  V\,dl^{\bar{\tilde{ \alpha}}}\,$, etc.},
\begin{eqnarray}
2 \,B^+ =  i\, V_l\,dl-i\,V_{\bar l}\,d\bar l - i\,V_r\,dr+i\, V_{\bar r}\,d\bar r+i\,V_w\,dw -
i\,V_{\bar w}\,d\bar w -i\,V_z\,dz +i V_{\bar z}\,d\bar z ,
\end{eqnarray}
\begin{eqnarray}
 2 \,B^- =  -i\, V_l\,dl+i\,V_{\bar l}\,d\bar l - i\,V_r\,dr+i\, V_{\bar r}\,d\bar r
 -i\,V_w\,dw + i\,V_{\bar w}\,d\bar w -i\,V_z\,dz +i V_{\bar z}\,d\bar z .
\end{eqnarray}
One notices that $ \Omega^\pm$ are not uniquely determined. Indeed we get that,
\begin{eqnarray}
\Omega^\pm\simeq \Omega^\pm+d\xi^\pm\,J_\pm,
\end{eqnarray}
where $d\xi^+$ and $d\xi^-$ are $(2,0)+(0,2)$ forms w.r.t. $J_+$ and $J_-$ resp. Examples of this 
are $\xi^+=V_l\,dl+V_{\bar l}\, d\bar l$,  $\xi^+=i\,V_l\,dl-i\,V_{\bar l}\, d\bar l$,  
$\xi^-=V_r\,dr+V_{\bar r}\, d\bar r$ and  $\xi^-=i\,V_r\,dr-i\,V_{\bar r}\, d\bar r$. We however will
use\footnote{These are the expressions introduced in \cite{Sevrin:2009na}. In \cite{Hull:2008vw} 
and \cite{Hull:2010sn} a different choice is being used.} the expressions in eqs.~(\ref{omplus}) and 
(\ref{ommin}) as they are such that when no chiral 
fields are present we can rewrite $ \Omega^+$ as,
\begin{eqnarray}
  \Omega^+ (X,Y)= 2\,g\big(X, (J_+-J_-)^{-1}Y\big).\label{opglob}
\end{eqnarray}
Similarly, when there are no twisted chiral fields one gets,
\begin{eqnarray}
  \Omega^- (X,Y)= 2\,g\big(X, (J_++J_-)^{-1}Y\big).\label{omglob}
\end{eqnarray}
Note that these expressions only exist in regular neighborhoods, at loci where type changing occurs they might not exist.
Using the previous expressions one also finds,
\begin{eqnarray}
b_--b_+=\frac 1 2 \,d\big(
-V_l\,dl-V_{\bar l}\,d\bar l
+V_r\,dr+V_{\bar r}\,d\bar r
-V_w\,dw-V_{\bar w}\,d\bar w
+V_z\,dz+V_{\bar z}\,d\bar z
\big).
\end{eqnarray}

We now turn back to generalized K\"ahler geometry. It is clear that the type of $ {\cal J}_+$ and 
$ {\cal J}_-$ is now given by $(k_+,k_-)=(n_c,n_t)$. With the expressions given above we have 
almost all ingredients to write down explicit expressions for the pure spinors 
associated with $ {\cal J}_+$ and $ {\cal J}_-$. Rests us to specify the relation between $b$ 
appearing in eqs.~(\ref{gcs22}) and (\ref{eigenbundle}) and $b_+$ and $b_-$ appearing in 
eq.~(\ref{gminb}). We choose,
\begin{eqnarray}
b&=&b_++\frac 1 2 \,d\big(-V_l\,dl-V_{\bar l}\,d\bar l +V_z\,dz+V_{\bar z}d\bar z\big)
\nonumber\\
&=&b_-+\frac 1 2 \,d\big(-V_r\,dr-V_{\bar r}\,d\bar r +V_w\,dw+V_{\bar w}d\bar w\big).
\label{bgauge}
\end{eqnarray}
Using the above one verifies along the lines of \cite{Hull:2010sn}
 that the pure spinors, eqs.~(\ref{psdef}), are given by,
\begin{eqnarray}
\phi_+&=&d\bar z^1\wedge d\bar z^2\wedge \cdots\wedge  d\bar z^{n_c}\wedge
e^{i\, \Omega^++\Xi^+}\,, \nonumber\\
\phi_-&=&d\bar w^1\wedge d\bar w^2\wedge \cdots\wedge  d\bar w^{n_t}\wedge
e^{i\, \Omega^-+\Xi^-}\,,\label{psexpl}
\end{eqnarray} 
where $\Omega^\pm$ are given in eqs.~(\ref{omplus}) and (\ref{ommin}) and $\Xi^\pm$ are
given by,
\begin{eqnarray}
\Xi^+&=&\frac 1 2 \,d\big(V_l\,dl+V_{\bar l}\,d\bar l
-V_z\,dz-V_{\bar z}\,d\bar z\big)\,, \nonumber\\
\Xi^-&=&\frac 1 2 \,d\big(V_r\,dr+V_{\bar r}\,d\bar r
-V_w\,dw-V_{\bar w}\,d\bar w\big)\, .\label{xidef}
\end{eqnarray} 
The expressions for the pure spinors appear differently from those in \cite{Hull:2010sn} but
they are actually the same (except for complex conjugation). The pure spinors $\phi_+$ and $\phi_-$ can 
still be multiplied by a function $f(\bar z)$ and $g(\bar w)$
resp.~ (which play an important role in
the analysis of the transformation properties of the generalized Calabi-Yau conditions) without changing 
the fact that they are closed but we reabsorbed 
them here by a holomorphic redefinition of the chiral and twisted chiral coordinates. Note that when the 
model is parametrized purely by twisted chiral superfields we find that $i\, \Omega^+ + \Xi^+ = - (i\, 
\Omega^- + \Xi^-)$, while a model parametrized by purely chiral superfields is characterized by $i\, 
\Omega^+ + \Xi^+ =  i\, \Omega^- + \Xi^-$.

Using these explicit expressions one verifies that,
\begin{eqnarray}
\big(\phi_+,\bar \phi_+\big)&=&(-1)^{n_c(n_c+1)/2+n_t+n_s}\,2^{n_t+2n_s}\,\det N_+\, \nonumber\\
\big(\phi_-,\bar \phi_-\big)&=&(-1)^{n_t(n_t+1)/2}\,2^{n_c+2n_s}\,\det N_-\, ,
\label{mukaip}
\end{eqnarray}
where in an obvious notation,
\begin{eqnarray}
N_+=\left(
\begin{array}{ccc}
V_{l\bar l}&V_{lr}&V_{l\bar w}\\
V_{\bar r\bar l}&V_{\bar r r}&V_{\bar r \bar w}\\
V_{w\bar l}&V_{wr}&V_{w\bar w}\end{array}
\right),
\end{eqnarray}
and,
\begin{eqnarray}
N_-=\left(
\begin{array}{ccc}
V_{l\bar l}&V_{l\bar r}&V_{l\bar z}\\
V_{r\bar l}&V_{r\bar r }&V_{ r \bar z}\\
V_{z\bar l}&V_{z\bar r}&V_{z\bar z}\end{array}
\right).
\end{eqnarray}
Viewing the result of the Mukai pairing as a $4n_s+2n_t+2n_c$-form, one should multiply the expressions
in eq.~(\ref{mukaip}) with $dL\wedge d\bar L\wedge dR\wedge d\bar R\wedge dW\wedge d\bar 
W\wedge dZ\wedge d\bar Z$ where $dL$ stands for $dl^1\wedge dl^2\wedge\cdots\wedge dl^{n_t}$ 
etc. The pure spinors in eq.~(\ref{psexpl}) are manifestly closed and using eq.~(\ref{mukaip}) 
the generalized Calabi-Yau condition eq.~(\ref{gcycd}) becomes,
\begin{eqnarray}
\frac{\det\big(N_+\big)}{\det\big(N_-\big)}\,=\mbox{ constant }\neq 0,\label{CYcd1}
\end{eqnarray}
fully consistent with the results in \cite{Hull:2010sn}. This has to be contrasted with the results obtained
in \cite{Grisaru:1997pg} where the one-loop counterterm for a general non-linear $\sigma$-model in $N=(2,2)$ superspace was calculated. The result is given by,
\begin{eqnarray}
{\cal S}_{1-\mbox{loop}}\,\propto\frac{1}{\varepsilon}\,\int d^2 \sigma \,d^2 \theta \, d^2 \hat \theta \,\ln\,\frac{\det\big(N_+\big)}{\det\big(N_-\big)}\,.
\end{eqnarray}
This vanishes provided,
\begin{eqnarray}
\frac{\det\big(N_+\big)}{\det\big(N_-\big)}\,=\pm|f_+(l,w,z)|^2
|f_-(r, \bar{w},z)|^2
,\label{CYcd2}
\end{eqnarray}
for some functions $f_+$ and $f_-$. This is clearly 
a weaker condition than the generalized Calabi-Yau condition in 
eq.~(\ref{CYcd1})! While eq.~(\ref{CYcd2}) guarantees that the generalized K\"ahler geometry is $N=(2,2)$ superconformal at the quantum level, the stronger condition eq.~(\ref{CYcd1}) has to be satisfied in order that the generalized K\"ahler geometry provides a solution to the supergravity equations of motion.   We will reconsider this in section 4 using concrete examples.

%:here

Let us conclude this section with a few interesting observations. As shown above we can (locally) define the two-forms $\Omega^+$, $\Omega^-$, once we know the metric $g$ and the torsion two-form potential $b$, see e.g.~expressions in eq.~(\ref{gminb}). Then constructing the pure spinors associated with $g$ and $b$ requires us to introduce the two-forms $\Xi^+$ and $\Xi^-$ as in eqs.~(\ref{xidef}). Now, we can turn the reasoning upside down. Consider the two pure spinors given in eqs.~(\ref{psexpl}) we can compute (locally) the metric and the torsion two-form potential once we know the two-forms $\Omega^+$, $\Omega^-$, $\Xi^+$ and $\Xi^-$:
\begin{eqnarray}
g = \frac{1}{2} \left( \Omega^+ J_+ + \Omega^- J_- + \Xi^+ - \Xi^- \right),
\end{eqnarray}
and,
\begin{eqnarray}
b =\frac{1}{2} \left( \Omega^- J_- - \Omega^+ J_+  - \Xi^+ - \Xi^- \right).
\end{eqnarray}

Consider now the particular case where $\ker
[J_+,J_-]= 0$, {\em i.e.} the $\sigma $-model is solely described
in terms of semi-chiral coordinates, then further simplifications occur \cite{Sevrin:1996jr}. We introduce
another closed two-form,
\begin{eqnarray}
  \Omega (X,Y)= 4\,g\big(X, [J_+,J_-]^{-1}Y\big),\label{globa1}
\end{eqnarray}
which is explicitly given by,
\begin{eqnarray}
\Omega =
 \left(\begin{array}{cc}
   0 & M_{lr}\\
   - M_{rl} &0
 \end{array}   \right),\label{bigomega}
\end{eqnarray}
where we labeled rows and columns in the order $(l, \bar l, r, \bar r)$.
One gets that $\Omega $ is a $(2,0)+(0,2)$ two-form with respect to both
$J_+$ and $J_-$,
\begin{eqnarray}
\Omega (J_+X,J_+Y)=-\Omega (X,Y),\qquad \Omega (J_-X,J_-Y)=-\Omega (X,Y),
\end{eqnarray}
and we also have that,
\begin{eqnarray}
\Omega=-2\Xi^+=2\Xi^-,
\end{eqnarray}
where $\Xi^\pm$ were defined in eq.~(\ref{xidef}). 
The relation of $\Omega $ with the previously introduced two-forms $
\Omega ^{+}$ and $ \Omega ^{-}$ is simple,
\begin{eqnarray}
  2\,\Omega \big(X,(J_++J_-)Y\big)=-\Omega ^+\big(X,Y\big),\qquad
  2\,\Omega \big(X,(J_+-J_-)Y\big)=+\Omega ^-\big(X,Y\big).
\end{eqnarray}
In terms of $\Omega $ we get particularly simple expressions for the
metric $g$ and the Kalb-Ramond two-form $b$,
\begin{eqnarray}
 g(X,Y)&=&\frac 1 4 \, \Omega \big(X,[J_+,J_-]Y\big),\nonumber\\
b(X,Y)&=&\frac 1 4 \, \Omega \big(X,\{J_+,J_-\}Y\big).\label{globa2}
\end{eqnarray}
Note that $\Omega $, and as a consequence $b$ as well, are both globally
well defined. This is indeed so modulo loci where type changing occurs
\cite{Gualtieri:2003dx}.

It is interesting to note here that while chiral (or twisted chiral)
coordinates are determined modulo a coordinate transformation involving
only chiral (or twisted chiral) coordinates such that the redefined
fields are still chiral (or twisted chiral), the parametrization freedom
becomes significantly larger once semi-chiral coordinates are present as
well. Indeed, in that case coordinate transformations of the form
$l\rightarrow l'(l,z,w)$ and $r\rightarrow r'(r,z,\bar w)$ are still
compatible with the constraints. Once semi-chiral fields enter the
picture the ambiguity on the exact form of the generalized K\"ahler
potential becomes also much larger than what is stated in
eq.~(\ref{genKahltrsf1}). Consider {\em e.g.} the case where only
semi-chiral fields are present. Making the coordinate transformation
\cite{Sevrin:1996jr},
\begin{eqnarray}
 q^{\tilde \alpha }=l^{\tilde \alpha },\qquad p^{\tilde \alpha }=V_{\tilde \alpha },
\end{eqnarray}
diagonalizes $J_+$ and the closed two-form $\Omega (X,Y)=4\, g(X,
[J_+,J_-]^{-1}Y)$ (see eq.~(\ref{bigomega})) reduces to its canonical
form,
\begin{eqnarray}
 \Omega = \left(
\begin{array}{cc}
 0& {\bf 1} \\
 -{\bf 1}& 0
\end{array}
 \right).
\end{eqnarray}
The coordinate transformation,
\begin{eqnarray}
 q^{\tilde\mu}=V_{\tilde \mu},\qquad p^{\tilde \mu}=r^{\tilde\mu},
\end{eqnarray}
diagonalizes $J_-$ and again the two-form $\Omega $ reduces to its
canonical form. It is clear that the coordinate transformations in either
of the two coordinate systems, which can be made without altering the
structure, are holomorphic with respect to the diagonalized complex
structure and canonical with respect to the Poisson structure $\Omega ^{-1}$. A
little thought shows that this ambiguity gets reflected by the fact that
the potential is only determined modulo a Legendre transform
\cite{Grisaru:1997ep}. I.e. both $\hat V$ and $V$,
\begin{eqnarray}
 \hat V (\hat l, {\bar{\hat l}},\hat r, {\bar{\hat r}})=
 V(l,\bar l,r,\bar r)-F(l,\hat l)-\bar F(\bar l, {\bar{\hat l}})
+G(r,\hat r)+\bar G(\bar r, {\bar{\hat r}}),\label{genKahtr}
\end{eqnarray}
describe exactly the same semi-chiral geometry. Again this fits well into the interpretation of the 
potential 
as the generating function of a canonical transformation \cite{Lindstrom:2005zr} which is indeed only 
defined up to a Legendre transformation.

The previous discussion also gives us a systematic way to determine the dependence of the
generalized K\"ahler potential on the semi-chiral fields. Indeed starting from the complex structures in 
some coordinates one first determines the coordinates which diagonalize $J_+$ and bring the 
(possibly degenerate) Poisson structure $[J_+,J_-]g^{-1}$ in its canonical form. Subsequently one 
finds the coordinates which do the same for $J_-$. Playing around with holomorphic canonical 
coordinate transformations in either coordinate system will usually lead to a relatively simple set of 
first order differential equations for the generalized K\"ahler potential. 

Finally, one verifies that the transformation,
\begin{eqnarray}
 V(l,\bar l,r,\bar r,w,\bar w, z, \bar z)\rightarrow
-V(l,\bar l,\bar r, r,z,\bar z, w, \bar w),\label{mirrorpot}
\end{eqnarray}
maps $\{g,H,J_+,J_-\}$ to $\{g,H,J_+,-J_-\}$. So it can be viewed as a local realization of mirror symmetry. Knowing how mirror symmetry is realized locally also allows us to investigate how it effects the pure spinors (\ref{psexpl}). Namely, under a mirror symmetry $\phi_+$ and $\phi_-$ are interchanged up to a $b$-transform,
\begin{eqnarray}
\phi_+ \longrightarrow e^{-b^{\rm m}} \wedge \phi_-, \\
\phi_- \longrightarrow e^{-b^{\rm m}} \wedge \phi_+,
\end{eqnarray}
with $b^m$ given by,
\begin{eqnarray}
b^{\rm m} = \frac{1}{2} d\left( V_l \, dl + V_{\bar l}\, d\bar l + V_r\, dr + V_{\bar r}\, d \bar r  \right). 
\end{eqnarray} 
Note that when there are only semi-chiral superfields (or no semi-chiral superfields at all) to parametrize the model, $b^{\rm m}$ is zero.

\section{Supersymmetric WZW-models}
Given a reductive Lie group\footnote{A reductive Lie group is a Lie group for which the Lie algebra can be written as the sum of an abelian and a semi-simple Lie algebra.} $\cal G$ and $g\in{\cal G}$ a group element
in some representation with index $x$ (see the appendix). The $N=(1,1)$
WZW action in $N=(1,1)$ superspace is given by,
\begin{eqnarray}
 {\cal S}&=& -\frac{k}{\pi x}\,\int_\Sigma  d^2\sigma \,d^2\theta \,\mbox{Tr}\big(
D_+gg^{-1}D_-gg^{-1}
 \big)+\nonumber\\
&&\frac{k}{\pi x}\,\int_\Xi dt \,d^2\sigma \,d^2\theta \,\mbox{Tr}\big(\partial _tg g^{-1}\big\{
D_+g g^{-1},D_-gg^{-1}
\}\big),
\end{eqnarray}
where $k\in\IN$, $\Sigma $ is the worldsheet and $\Xi $ is a
three-dimensional manifold which has the worldsheet as its boundary,
$\partial \Xi=\Sigma$. We denoted the ``third coordinate'' on $\Sigma $
by $t$. The equations of motion read,
\begin{eqnarray}
 D_+\big(D_-gg^{-1}\big)=D_-\big(g^{-1}D_+g\big)=0.
\end{eqnarray}
Besides the $N=(1,1)$ superconformal invariance the WZW action is also
invariant under the affine transformations,
\begin{eqnarray}
 g\rightarrow h_-\,g\,h_+,\label{wzwiso}
\end{eqnarray}
where $h_\pm\in{\cal G}$ and,
\begin{eqnarray}
 D_+h_-=D_-h_+=0.\label{wzwiso1}
\end{eqnarray}

In \cite{Spindel:1988nh} it was shown that the model has an $N=(2,2)$
supersymmetry provided $\cal G$ is an even-dimensional reductive Lie
group. The extra supersymmetry transformations can be written as,
\begin{eqnarray}
 \big(g^{-1}\delta g\big)^A=\varepsilon^+\IJ_+^A{}_B\big(g^{-1}D_+ g\big)^B+
 \varepsilon ^-\big(LR^{-1}\big)^A{}_C\IJ_-^C{}_D\big(RL^{-1}\big)^D{}_B
 \big(g^{-1}D_- g\big)^B,
\end{eqnarray}
or,
\begin{eqnarray}
 \big(\delta g g^{-1}\big)^A=
\varepsilon ^+\big(RL^{-1}\big)^A{}_C\IJ_+^C{}_D\big(LR^{-1}\big)^D{}_B
 \big(D_+ gg^{-1}\big)^B+
 \varepsilon^-\IJ_-^A{}_B\big(D_- gg^{-1}\big)^B,
\end{eqnarray}
where the complex structures on the Lie algebra, $\IJ_\pm^A{}_B$, are
related to those on the Lie group, $J_\pm^a{}_b$, by,
\begin{eqnarray}
 \IJ_+^A{}_B=L_c^A\,J_+^c{}_d\,L^d_B,\qquad
\IJ_-^A{}_B=R_c^A\,J_-^c{}_d\,R^d_B,
\end{eqnarray}
where $L_a^B$ and $R_a^B$ are the left and right invariant vielbeins (see
the appendix for our conventions). Note that $(LR^{-1})^A{}_B$ is a group transformation in the adjoint representation. The conditions for $N=(2,2)$
supersymmetry (as given at the beginning of section 2) can be
reformulated as conditions on the Lie algebra \cite{Spindel:1988nh}:
\begin{enumerate}
  \item $\IJ_\pm^A{}_B$ are constant and satisfy $\IJ_\pm^A{}_C\,\IJ_\pm^C{}_B=-\delta
  _B^A\,$,
  \item $\IJ_\pm^C{}_A\,\IJ_\pm^D{}_B\,\eta_{CD}=\eta_{AB}\,$,
  \item $f_{DEA}\,\IJ_\pm^D{}_B\, \IJ_\pm^E{}_C+f_{DEB}\,\IJ_\pm^D{}_C\, \IJ_\pm^E{}_A+f_{DEC}\,\IJ_\pm^D{}_A\,
  \IJ_\pm^E{}_B=f_{ABC}$.
\end{enumerate}
In a basis where a complex structure is diagonal such that it has eigenvalue $+i$ on $T_A$
and eigenvalue $-i$ on $T_{\bar A}$, the second condition simply states that
$\eta_{AB}= \eta_{\bar A\bar B}=0$ and the last condition becomes equivalent to 
$f_{ABC}=f_{\bar A\bar B\bar C}=0$.

These conditions were solved in \cite{Spindel:1988nh}. There it was shown
that a complex structure on the Lie algebra is almost fully characterized
by a Cartan decomposition where the complex structures are diagonal on
the generators corresponding with positive (negative) roots with
eigenvalue $+i$ ($-i$). The complex structure maps the Cartan subalgebra
(CSA) to itself such that the Cartan-Killing metric is hermitian. As any two Cartan decompositions of the 
Lie algebra are  related through group conjugation, the remaining freedom lies in the choice of the 
complex structure on the Lie algebra.

Consider now the transformation in eq.~(\ref{wzwiso}) with $h_+$ and $h_-$ constant. Under this the complex structures transform as,
\begin{eqnarray}
\IJ_+^A{}_B&\rightarrow& \IJ'{}_+^A{}_B=\big(
h_+^{-1}\IJ_+ h_+
\big)^A{}_B, \nonumber\\
\IJ_-^A{}_B&\rightarrow& \IJ'{}_-^A{}_B=\big(
h_-\IJ_- h_-^{-1}
\big)^A{}_B,\label{csiso}
\end{eqnarray}
where $h_+$ and $h_-$ are now in the adjoint representation. 
Only the isometries which leave the complex structures invariant are compatible with the $N=(2,2)$ 
supersymmetry. 

We focus now on one of the two complex structures on the Lie algebra which we will call $\IJ$. Using 
eq.~(\ref{csiso}) one finds that invariance of $\IJ$ under an infinitesimal isometry generated by 
$T_A$ requires,
\begin{eqnarray}
f_{AD}{}^C\,\IJ^D{}_B=\IJ^C{}_D\,f_{AB}{}^D.\label{isoJ}
\end{eqnarray}
This condition is easily analyzed by choosing a complex basis for the Lie algebra such that $\IJ$ is 
diagonal. One finds that eq.~(\ref{isoJ}) is satisfied provided $T_A$ belongs to the Cartan sub 
algebra and it is immediate that $\IJ'=\IJ$ in eq.~(\ref{csiso}) iff.~$h_+$ and $h_-$ are in the 
maximal torus of the group. 
This was to be expected as precisely the group elements which belong to the maximal torus of the 
group leave the Cartan decomposition invariant. 
The previous discussion holds for constant isometries. In order to determine 
the affine extensions of it one needs more information about the superfield content of the particular model 
under study as will be illustrated in explicit examples later on.

Concluding:  a manifest $N=(2,2)$ supersymmetric formulation of 
WZW-models will not be invariant under the full affine symmetry group, eqs.~(\ref{wzwiso}) and 
(\ref{wzwiso1}) but only under a maximal abelian subset of those. The fact that generically the 
commutator of an ($N=(1,1)$) isometry with the second supersymmetry yields a ``new'' 
supersymmetry was already noted in \cite{Hull:1990qh} and \cite{Hull:1990qf} where the 
author(s) propose that the action of such isometries should be paired with a simultaneous 
transformation of the complex structure. From the previous discussion this is now clear. Performing a group 
transformation eq.~(\ref{wzwiso}) which does not belong to the maximal torus brings one to 
another Cartan decomposition of the Lie algebra implying a (continuous) movement in the moduli space of 
the complex structure. 

Let us now turn to the superfield content of a particular WZW-model. Once a choice for $\IJ_+$ 
and $\IJ_-$ on the Lie algebra is made, the field content is completely fixed. 
In order that no semi-chiral fields are needed,
\begin{eqnarray}
[\IJ_+,G\,\IJ_-\,G^{-1}]=0,\label{wzwcom}
\end{eqnarray}
should hold where $G$ is an arbitrary element of the group in the adjoint representation. Writing $G=e^{i\, \alpha}$ and analyzing 
eq.~(\ref{wzwcom}) through first order in $\alpha$ shows that this is only possible on 
$SU(2)\times U(1)$ \cite{Rocek:1991vk}. All other groups will necessarily require semi-chiral superfields.

In general the number of chiral (twisted chiral resp.) directions is given by the dimension of $\ker(\IJ_+-
G\,\IJ_-G^{-1})$ ($\ker(\IJ_++G\,\IJ_-G^{-1})$ resp.), with once again $G$ an arbitrary group element in 
the adjoint representation. The remaining directions will then be parameterized by semi-chiral super 
fields. 

An obvious choice is given by taking both complex structures on the Lie algebra to be equal, {\em i.e.} 
$\IJ_-=\IJ_+$. In this case $\ker(\IJ_++G\,\IJ_-G^{-1})$ is always zero and $\ker(\IJ_+-G\,\IJ_-G^{-1})$
can easily be analyzed through first non-trivial order in the group parameters. An explicit check of this for 
the non-abelian reductive rank 2 Lie groups is given in table \ref{Jalg1}. From these examples one
expects that the choice $\IJ_-=\IJ_+$ maximizes the number of semi-chiral fields
needed to parameterize the model.
\begin{table}
\begin{center}
\begin{tabular}{|c|c|c|}
\hline 
Group& $n_s$& $n_c$\\
\hline\hline
$SU(2)\times U(1)$&1&0 \\ 
$SU(2)\times SU(2)$&1&1 \\ 
$SU(3)$&2&0 \\ 
$SO(5)$&2&1 \\ 
$G_2$&3&1 \\ 
\hline
\end{tabular}
\caption{The superfield content for the rank 2 non-abelian reductive Lie groups when
taking $\IJ_-=\IJ_+$ on the Lie algebra. The number of semi-chiral and chiral superfields resp.~are
denoted by $n_s$ and $n_c$ resp.\label{Jalg1}}
\end{center}
\end{table}

Another obvious choice for the complex structures on the Lie algebra is given by taking $\IJ_+$ and 
$\IJ_-$ to be equal on the roots but having opposite sign on the CSA, {\em i.e.} 
$\IJ_-|_{\mbox{\small roots}}=\IJ_+|_{\mbox{\small roots}}$ and $\IJ_-|_{\mbox{\small CSA}}=-
\IJ_+|_{\mbox{\small CSA}}$. 
Analyzing both $\ker(\IJ_++G\,\IJ_-G^{-1})$ and $\ker(\IJ_+-G\,\IJ_-G^{-1})$ through forst non-trivial
order gives the superfield content which for rank 2 non-abelian reductive group manifolds is given
in table \ref{Jalg2}. So here we anticipate that this choice minimizes the number of semi-chiral 
superfields.

A systematic analysis of the superfield content as a function of the choices made for the complex 
structures on the Lie algebra seems certainly feasible and is left to future investigation.
\begin{table}[h]
\begin{center}
\begin{tabular}{|c|c|c|c|}
\hline 
Group& $n_s$& $n_c$&$n_t$\\
\hline\hline
$SU(2)\times U(1)$&0&1&1 \\ 
$SU(2)\times SU(2)$&1&0 &1\\ 
$SU(3)$&1&1& 1\\ 
$SO(5)$&1&1 &2\\ 
$G_2$&2&1 &2\\ 
\hline
\end{tabular}
\caption{The superfield content for the rank 2 non-abelian reductive Lie groups when
taking $\IJ_-|_{\mbox{\small roots}}=\IJ_+|_{\mbox{\small roots}}$ and $\IJ_-|_{\mbox{\small CSA}}=-
\IJ_+|_{\mbox{\small CSA}}$ on the Lie algebra. The number of semi-chiral, chiral and 
twisted chiral superfields 
resp.~are denoted by $n_s$, $n_c$ and $n_t$ resp.\label{Jalg2}}
\end{center}
\end{table}

\section{$SU(2)\times U(1)$}
The simplest non-trivial $N=(2,2)$ WZW-model is the one on $SU(2)\times U(1)$ (or the Hopf 
surface $S^3\times S^1$). Its formulation in terms of one chiral and one twisted chiral superfield was 
introduced in \cite{Rocek:1991vk}. An alternative formulation in terms of a semi-chiral multiplet was 
presented in \cite{Sevrin:1996jr} (see also the discussion in \cite{Ivanov:1994ec}). In this section we will 
systematically analyze the possible choices one can make for the left and right complex structures and 
identify the superfields and generalized K\"ahler potential.  
\subsection{The complex structures on the Lie algebra}
We denote the three Pauli matrices by $\sigma _j$ with $j\in\{1,2,3\}$
and the $2\times 2$ unit matrix by $\sigma _0$. We take the generators of
the Lie algebra of $SU(2)\times U(1)$ in the fundamental representation
as,
\begin{eqnarray}
 h=\frac 1 2 \big(\sigma _3+i\,\sigma _0\big),\quad
 \bar h=\frac 1 2 \big(\sigma _3-i\,\sigma _0\big), \quad
 e=\frac 1 2 \big(\sigma _1+i\,\sigma _2\big),\quad
\bar  e=\frac 1 2 \big(\sigma _1-i\,\sigma _2\big),
\end{eqnarray}
where $e$ ($\bar e$) corresponds to the positive (negative) root.
Labelling rows and columns in the order $(he\bar h\bar e)$, we get so two
possibilities for the complex structure on the Lie algebra,
\begin{eqnarray}
\IJ_1 = \left(\begin{array}{cc} i \sigma_0 & 0 \\ 0 & - i \sigma_0
\end{array} \right), \qquad
\IJ_2 = \left(\begin{array}{cc} - i \sigma_3 & 0 \\ 0 & i \sigma_3
\end{array} \right),
\end{eqnarray}
reflecting the two possible choices we can make for the complex structure
on the CSA. Now we still have to account for the freedom in choosing a
Cartan decomposition. As two Cartan decompositions are related by a group
conjugation, we get the general expression for the complex structure by
taking $G \IJ G^{-1}$ where $G$ is a group element in
the adjoint representation. In this way we get,
\begin{eqnarray}
\IJ_1(\theta ,\phi ) &=& i\,\left(\begin{array}{cc} \cos\theta \, \sigma_0 &
\sin\theta \,e^{-i\phi }\,\sigma _2 \\
\sin\theta \,e^{i\phi }\,\sigma _2 & - \cos\theta \, \sigma_0
\end{array} \right), \nonumber\\
\IJ_2(\theta ,\phi ) &=& i\,\left(\begin{array}{cc} - \cos\theta \, \sigma_3 +
\sin\theta\big(\sin\phi \,\sigma _1+\cos\phi \,\sigma _2\big)& 0 \\
0 &   \cos\theta \, \sigma_3 -\sin\theta \big(\sin\phi \,\sigma _1-
\cos\phi \,\sigma _2\big)
\end{array} \right),\nonumber\\
\end{eqnarray}
where $\phi \in [0,2\pi ]$ and $\theta \in [0,\pi ]$. Note that we could
have obtained the above expressions in a different way. As $SU(2)\times
U(1)$ allows for an $N=(4,4)$ supersymmetry \cite{Spindel:1988nh} we can
find an additional one-parameter family of complex structures $\IK_1(\phi
)$ and $\IK_2(\phi )$ such that $\{\IJ_1,\IK_1(\phi )\}=\{\IJ_2,\IK_2(\phi )\}=0$.
Using this one immediately gets $\IJ_1(\theta ,\phi )=\cos\theta\,
\IJ_1+\sin\theta\, \IK_1(\phi )$ and $\IJ_2(\theta ,\phi )=\cos\theta
\,\IJ_2+\sin\theta \,\IK_2(\phi )$.

\subsection{The complex geometry of the group}\label{subsection:geogroup}
We parameterize a group element $g$ in the fundamental representation as,
\begin{eqnarray}
g= e^{i\rho }\,\left(
\begin{array}{cc} \cos \psi \, e^{i\varphi _1} & \sin \psi \,e^{i\varphi _2}\\
-\sin\psi\, e^{-i\varphi _2} & \cos\psi\, e^{-i\varphi _1}\end{array}\right),
\label{su2u1coor}
\end{eqnarray}
where $\varphi _1,\, \varphi_2,\, \rho \in \IR\,\mbox{mod}\,2 \pi $ and
$\psi\in[0, \pi /2]$. So locally this has the topology of a line segment parameterized
by $\psi$ times a $T^3$ parameterized by $\rho$, $ \varphi_1$ and $\varphi_2$.
At the endpoints of the line, $\psi=0$ or $\psi=\pi/2$ resp., the $T^3$ degenerates to a
$T^2$ parameterized by $\rho$ and $\varphi_1$ or $\rho$ and $\varphi_2$ resp. The group
manifold $S^3\times S^1$ can also be viewed as a rational Hopf surface defined by 
$\big(\IC^2\backslash(0,0)\big)/\Gamma$ where elements of $\Gamma$ act on 
$(w,z)\in\IC^2\backslash(0,0)$ as $(w,z) \rightarrow e^{2 \pi n}(w,z)$, $n\in\IZ$.
This becomes manifest when identifying,
\begin{eqnarray}
w&=&\cos\psi\,e^{-\rho-i \varphi_1}, \nonumber\\
z&=&\sin\psi\,e^{-\rho+i \varphi_2}\,,
\end{eqnarray}
where $\psi$, $\rho$, $\varphi_1$ and $\varphi_2$ were introduced above.

With this the metric is given by,
\begin{eqnarray}
 ds^2=\frac{k}{2\pi }\,\big( d\rho ^2+d\psi ^2+\cos^2\psi\, d\varphi_1 ^2+\sin^2\psi\, d \varphi _2^2\big),
\end{eqnarray}
and the 3-form is,
\begin{eqnarray}
H=\frac{k}{2\pi }\,\sin 2\psi\,
 d\varphi _1\wedge d\varphi _2\wedge d\psi  .
\end{eqnarray}
We can choose a gauge such that $b_{\varphi_1\varphi _2}=-k\,\cos2\psi/4\pi $ and
all other components zero.

As we can choose $J_-$ independently of $J_+$ we get four different choices for the complex structures,
\begin{enumerate}
  \item $J_+^a{}_b=L^a_C\,\IJ_1(\theta ,\phi )^C{}_D\,L^D_b$,
  $J_-^a{}_b=R^a_C\,\IJ_2(\theta' ,\phi ' )^C{}_D\,R^D_b\,$.
  \item $J_+^a{}_b=L^a_C\,\IJ_2(\theta ,\phi )^C{}_D\,L^D_b$,
  $J_-^a{}_b=R^a_C\,\IJ_1(\theta ',\phi' )^C{}_D\,R^D_b\,$.
  \item $J_+^a{}_b=L^a_C\,\IJ_1(\theta ,\phi )^C{}_D\,L^D_b$,
  $J_-^a{}_b=R^a_C\,\IJ_1(\theta ',\phi ' )^C{}_D\,R^D_b$.
  \item $J_+^a{}_b=L^a_C\,\IJ_2(\theta ,\phi )^C{}_D\,L^D_b$,
  $J_-^a{}_b=R^a_C\,\IJ_2(\theta' ,\phi' )^C{}_D\,R^D_b$.
\end{enumerate}
The first and second choices give commuting complex structures where both
$\ker (J_+-J_-)$ and $\ker (J_++J_-)$ are two-dimensional. So here we
will end up with a superspace description in terms of one chiral
superfield $z$ and one twisted chiral superfield $w$. For the last two
choices we find that almost everywhere\footnote{At certain points type changing occurs, more on this in section
\ref{section:432}.} $\ker [J_+,J_-]= 0$ holds and we expect a
superspace description in
terms of a semi-chiral multiplet $(l,r)$. We can explicitly write these
coordinates as a function of the coordinates on the group manifold we
previously introduced in eq.~(\ref{su2u1coor}).\\
\underline{\em i. The first choice}\\
 For the first choice we start by introducing,
\begin{eqnarray}
 \tilde w=\cos\psi \,e^{-\rho -i\,\varphi _1},\qquad
 \tilde z=\sin\psi \,e^{-\rho +i\,\varphi _2},\label{aux1}
\end{eqnarray}
which diagonalizes $J_+$ and $J_-$ in the case where $\theta =\phi
=\theta '=\phi '=0$. Making the coordinate transformation,
\begin{eqnarray}
\begin{array}{ll}
x_1 = - e^{i \phi} \cos \frac{ \theta}{2}\, \tilde{ w} +
i\, \sin \frac{\theta}{2}\, \bar{\tilde{z}}, & \bar x_1 =
- e^{-i \phi} \cos \frac{ \theta}{2}\, \bar {\tilde{ w}}
- i\, \sin \frac{\theta}{2}\, \tilde{z}, \\
& \\
x_2 =  e^{i \phi} \cos \frac{ \theta}{2}\, \tilde{ z} +
i\, \sin \frac{\theta}{2}\, \bar{\tilde{w}}, & \bar x_2 =
e^{-i \phi} \cos \frac{ \theta}{2}\, \bar {\tilde{ z} }
- i\, \sin \frac{\theta}{2}\, \tilde{w}\, ,
\end{array}
\end{eqnarray}
diagonalizes $J_+$ for generic values of $\theta $ and $\phi $. Making a
holomorphic transformation (w.r.t. $J_+$),
\begin{eqnarray}
\begin{array}{ll}
w = - \cos \frac{\theta'}{2} x_1 - i \sin \frac{\theta'}{2}\,
e^{i \phi'}  x_2, & \bar w = -\cos \frac{\theta'}{2} \bar{ x_1} + i \sin \frac{\theta'}{2}\, e^{-i \phi'} \bar{x_2}\, ,\\&\\
z =  i \sin \frac{ \theta'}{2}\,  e^{-i \phi'} x_1+ \cos
\frac{\theta'}{2} x_2,
& \bar z =  - i \sin \frac{ \theta'}{2}\,  e^{i \phi'} \bar{x_1} + \cos \frac{\theta'}{2} \bar {x_2}\, ,
\end{array}
\label{diagco}
\end{eqnarray}
diagonalizes $J_-$ for arbitrary values of $\theta '$ and $\phi '$ as
well. One finds the generalized K{\"a}hler potential,
\begin{eqnarray} V(z,
w, \bar z, \bar w) = \frac{k}{4\pi }\,\Big(\int^{\frac{z \bar z}{ w \bar w}} \frac{dq}{q}
\ln\left(1+q\right) - \frac{1}{2} \left( \ln w \bar w \right)^2 \Big)
,\label{su2u1ctc4moduli}
\end{eqnarray}
which correctly reproduces the metric,
\begin{eqnarray}
 ds^2=\frac{k}{2\pi }\,\frac{1}{z\bar z+w\bar w}\,\big(dz\,d\bar z+dw\,d\bar w\big),
\label{gchitwi}
\end{eqnarray}
and torsion,
\begin{eqnarray}
H&=&\frac{k}{4\pi }\,\Big(\frac{\bar w}{(z\bar z+w\bar w)^2}\,dz\wedge d\bar z\wedge dw-
\frac{w}{(z\bar z+w\bar w)^2}\,dz\wedge d\bar z\wedge d\bar w+\nonumber\\
&&\frac{\bar z}{(z\bar z+w\bar w)^2}\,dz\wedge dw\wedge d\bar w-
\frac{ z}{(z\bar z+w\bar w)^2}\,d\bar z\wedge dw\wedge d\bar w\Big)\,,\label{Hchtwch}
\end{eqnarray}
in these coordinates.\\
\underline{\em ii. The second choice}\\
This case is almost identical to the first one except that our starting
point is now,
\begin{eqnarray}
 \tilde w=\cos\psi \,e^{+\rho -i\,\varphi _1},\qquad
 \tilde z=\sin\psi \,e^{+\rho +i\,\varphi _2},\label{aux2}
\end{eqnarray}
instead of eq.~(\ref{aux1}), so $\rho \rightarrow-\rho $. The resulting
generalized K{\"a}hler potential
is still given by eq.~(\ref{su2u1ctc4moduli}).\\
\underline{\em iii. The third choice}\\
We get that,
\begin{eqnarray}
 l=w,\quad \bar l=\bar w,\quad r=\frac{\bar w}{z},\quad \bar r=\frac{w}{\bar z},\label{semigroup}
\end{eqnarray}
where $z$ and $w$ are the expressions given in eq.~(\ref{diagco}). The
generalized K{\"a}hler potential is now,
\begin{eqnarray}
 V(l,\bar l,r,\bar r)=\frac{k}{4\pi }\Big(\ln \frac{l}{\bar r}\,\ln \frac{\bar l}{r}\,-\int^{r\bar r}\frac{dq}{q}\,
 \ln\big(1+q\big)\Big).\label{Vsemi}
\end{eqnarray}
Using this we calculate the metric,
\begin{eqnarray}
ds^2=\frac{k}{2\pi }\,\Big(\frac{1}{l\bar l}\,dl\,d\bar l+\frac{1}{r\bar r}\,\frac{1}{1+r\bar r}\,drd\bar r-
\frac{1}{lr}\,\frac{1}{1+r\bar r}\,dl\,dr-\frac{1}{\bar l\bar r}\,\frac{1}{1+r\bar r}\,
d\bar l\,d\bar r\Big),
\end{eqnarray}
and the torsion 3-form,
\begin{eqnarray}
H=\frac{k}{4\pi }\Big(\frac{1}{l}\,\frac{1}{(1+r\bar r)^2}\,dl\wedge dr\wedge d\bar r-
\frac{1}{\bar l}\,\frac{1}{(1+r\bar r)^2}\,  d\bar l\wedge dr\wedge d\bar r \Big).
\end{eqnarray}
The complex structures are,
\begin{eqnarray}
 J_+&=&\left(\begin{array}{cccc}
   +i & 0 & 0 & 0 \\
   0 &-i & 0 & 0 \\
   0 & -2i\,\frac{r}{\bar l} & +i&0\\
   +2i\,\frac{\bar r}{l}&0&0&-i
 \end{array}   \right),\nonumber\\
J_-&=&\left(\begin{array}{cccc}
   i&0&0&-2i\,\frac{l}{\bar r}\,\frac{1}{1+r\bar r} \\
   0& -i& +2i\,\frac{\bar l}{ r}\,\frac{1}{1+r\bar r}&0\\
   0 & 0 & +i & 0 \\
   0 & 0 & 0 & -i
 \end{array}   \right),\label{thecomstr}
\end{eqnarray}
where we labelled the rows and columns in the order $l\bar l r\bar r$.\\
\underline{\em iv. The fourth choice}\\
Here as well the semi-chiral coordinates are given by
eq.~(\ref{semigroup}) but now with $z$ and $w$ as in the second choice.
The generalized K{\"a}hler potential and all other expressions are then obviously the same
as for the third choice.

As mentioned before, the potential in eq.~(\ref{Vsemi}) is only determined modulo a Legendre 
formulation (\ref{genKahtr}).  For completeness we give here three alternative forms. In the first we 
Legendre 
transform with respect to $l$ and $\bar l$ which gives,
\begin{eqnarray}
V_1(l',\bar l', r, \bar r)&=&V(l,\bar l, r, \bar r)-
\frac{k}{4\pi }\Big(\ln l\,\ln l'+\ln \bar l \, \ln \bar l'\Big) \nonumber\\
&=&\frac{k}{4\pi }\Big(-\ln l'\ln \bar l'-\ln l'\ln \bar r-\ln \bar l'\ln r \nonumber\\
&&-\int^{r \bar r} \frac{dq}{q}\,\ln(1+q)\Big),
\end{eqnarray}
with,
\begin{eqnarray}
l'= \frac{\bar l}{r}\,, \qquad \bar l'= \frac{l}{\bar r}\,.
\end{eqnarray}
Performing a further coordinate transformation,
\begin{eqnarray}
r'=\frac 1 r \,,\qquad \bar r ' = \frac{1}{\bar r}\,,
\end{eqnarray}
results in the potential (modulo terms which can be removed by a generalized K\"ahler transformation),
\begin{eqnarray}
 V_2(l',\bar l',r',\bar r')=\frac{k}{4\pi }\Big(-\ln \frac{l'}{ r'}\,\ln \frac{\bar l'}{\bar r'}\,+\int^{r'\bar r'}\frac{dq}{q}\,
 \ln\big(1+q\big)\Big),\label{piano1}
\end{eqnarray}
which is recognized as the ``mirror transform'' of eq.~(\ref{Vsemi}). 

An alternative form is obtained by Legendre transforming eq.~(\ref{Vsemi}) with respect to $r$ 
and $\bar r$. This results in,
\begin{eqnarray}
V_3(l,\bar l, r', \bar r')&=&V(l,\bar l, r, \bar r)-\frac{k}{4\pi }\Big(\ln r\,\ln r'+\ln \bar r \, \ln \bar r'\Big) \nonumber\\
&=&\frac{k}{4\pi }\Big(-\ln r'\ln\bar r'-\ln l\ln\bar r '-\ln \bar l\ln r' \nonumber\\
&&+\int^{l\bar l r'\bar r'}
\frac{dq}{q}\,\ln\big(1\pm\sqrt{1-4q}\big)-\ln 2\, \ln(l\bar l r'\bar r')\Big)\,,\label{piano}
\end{eqnarray}
where,
\begin{eqnarray}
r'= \frac{\bar r}{l}\, \frac{1}{1+r\bar r}\,,
\end{eqnarray}
and the last term in eq.~(\ref{piano}) can be removed by a generalized K\"ahler transformation. 
The mirror transform of eq.~(\ref{piano}) is obtained by taking the Legendre transform of 
eq.~(\ref{piano1}) with respect to $r'$ and $\bar r'$,
\begin{eqnarray}
V_4(l,\bar l, r', \bar r')&=&V_2(l,\bar l, r, \bar r)+\frac{k}{4\pi }\Big(\ln r\,\ln r'+\ln \bar r \, \ln \bar r'\Big) \nonumber\\
&=&\frac{k}{4\pi }\Big(\ln r'\ln\bar r'+\ln l\ln r '+\ln \bar l\ln\bar r' \nonumber\\
&&-\int^{l\bar l r'\bar r'}
\frac{dq}{q}\,\ln\big(1\pm\sqrt{1-4q}\big)+\ln 2\, \ln(l\bar l r'\bar r')\Big)\,,\label{piano2}
\end{eqnarray}
where,
\begin{eqnarray}
r'= \frac{\bar r}{\bar l}\, \frac{1}{1+r\bar r}\,,
\end{eqnarray}
and once again we can remove the last term in the potential by a generalized K\"ahler transformation.

\subsection{Relating the semi-chiral and the chiral/twisted chiral formulations}
The simple relations in eq.~(\ref{semigroup}) suggest an equally simple
relation at the level of the $\sigma $-models. In \cite{Grisaru:1997ep}
it was shown that -- provided an adequate isometry is present -- a
semi-chiral multiplet is T-dual to a pair of superfields consisting of a
chiral and a twisted chiral superfield. In
\cite{Lindstrom:2007vc}-\cite{Merrell:2007sr} the underlying gauge
structure has been developed. In \cite{martinWIP} it was noted that
T-dualizing along the $S^1$ of $S^1\times S^3$ requires an isometry of the aforementioned type and 
therefore should map the chiral/twisted chiral description to the semi-chiral description and vice-versa.
We will use that observation here in a slightly different setting. Our
starting point is the potential in eq.~(\ref{Vsemi}). From the discussion
in section \ref{subsection:geogroup} and eq.~(\ref{semigroup}), it is
clear that in order to dualize along the $S^1$ of $S^3\times S^1$, the relevant isometry is given by,
\begin{eqnarray}
\delta l=\epsilon \,l,\qquad \delta\bar  l=\epsilon \,\bar l,\qquad
\delta r=0,\qquad \delta \bar r=0,\label{isob1}
\end{eqnarray}
where $\epsilon \in\IR$ and constant. The potential eq.~(\ref{Vsemi}) is
invariant modulo a generalized K{\"a}hler transformation ({\em i.e.} it is invariant modulo superspace 
total derivative terms). We generalize the approach developed in \cite{Hull:1985pq} and introduce an 
(auxiliary)
semi-chiral multiplet $\hat l$, $\bar{ \hat l}$, $\hat r$, $\bar{ \hat
r}$ which under the isometry transforms as,
\begin{eqnarray}
 \delta \hat l=\epsilon \,\ln l,\qquad \delta\bar{ \hat l}=\epsilon \,\ln\bar  l,
 \qquad\delta \hat r = \epsilon \,\ln r,\qquad \delta\bar{ \hat r} = \epsilon \,\ln \bar r.\label{isob2}
\end{eqnarray}
The potential,
\begin{eqnarray}
 V_0=\frac{k}{4\pi }\,\Big(\ln \frac{l}{\bar r}\,\ln \frac{\bar l}{r}\,-\int^{r\bar r}\frac{dq}{q}\,
 \ln\big(1+q\big)-\hat l-\bar{ \hat l}+\hat r+\bar{ \hat r}\Big),
\end{eqnarray}
is now exactly invariant under the isometry given by eqs.~(\ref{isob1}) and (\ref{isob2}). 
Adding the extra terms did
not change anything as they are superspace total derivative terms. We now
gauge the isometry. The gauge parameters $\epsilon _l$, $\bar \epsilon
_{\bar l}$, $\epsilon _r$ and $\bar \epsilon _{\bar r}$ form a
semi-chiral multiplet,
\begin{eqnarray}
 \bar \ID_+ \epsilon _l=\ID_+\bar \epsilon_{\bar l} = \bar \ID_-\epsilon _r=
 \ID_-\bar \epsilon_{\bar r} =0.
\end{eqnarray}
We introduce the unconstrained gauge fields $X\in\IR$ and $Y,\,\bar
Y\,(=Y^\dagger)\in\IC$. The potential,
\begin{eqnarray}
 V_1=V_0+\frac{k}{4\pi }\,\Big(i\,X\,\ln \frac{l}{\bar l}+\frac 1 2 \, X^2+Y\,\ln r+\bar Y\ln \bar r\Big),
\end{eqnarray}
is invariant under the gauge transformations,
\begin{eqnarray}
 &&\delta l=\epsilon _l\,l\,,\qquad \delta \bar l=\bar \epsilon _{\bar l}\,\bar l\,,
 \qquad \delta r=0\,,\qquad \delta \bar r=0\,,\nonumber\\
 &&\delta \hat l=\epsilon _l\,\ln l\,,\qquad \delta \bar{\hat l}=\bar \epsilon _{\bar l}\,\ln \bar l\,,
 \qquad \delta r=\epsilon _r\,\ln  r\,,\qquad \delta \bar{\hat r}=\bar\epsilon _{\bar r}\,\ln \bar r\,,\nonumber\\
 &&\delta X=-i\big(\epsilon _l-\bar \epsilon _{\bar l}\big)\,,
 \qquad \delta Y=\epsilon _l-\epsilon _r\,, \qquad \delta \bar Y=\bar \epsilon _{\bar l}-\bar \epsilon _{\bar r }\,.
\end{eqnarray}
The gauge invariant fieldstrengths are given by,
\begin{eqnarray}
&&  F=i\,\ID_+\bar \ID_-\big(X+i\,Y\big),\qquad \bar F=i\,\bar \ID_+\ID_-\big(X-i\,\bar Y\big),\nonumber\\
&& G=i\,\ID_+\ID_-\bar Y,\qquad \bar G=i\,\bar \ID_+\bar \ID_-Y\,.
\end{eqnarray}
Introducing the Lagrange multipliers $u,\,\bar u,\,v,\bar v$ which are
unconstrained complex superfields, we write the first order potential,
\begin{eqnarray}
 V_2=V_1+ \frac{k}{4\pi }\,\Big(u\,\bar F+\bar u\,F+ v\, \bar G+\bar v\,G\Big),\label{VV2}
\end{eqnarray}
which is equivalent to the original (ungauged) model. Indeed integrating
over the Lagrange multipliers $u$ and $v$ puts the fieldstrengths to zero and the
gauge fields can be gauged away. Upon integrating eq.~(\ref{VV2}) by
parts we get,
\begin{eqnarray}
 V_2=V_1-\frac{k}{4\pi }\,\Big(i\,X\,\ln\frac{w}{\bar w}\,+Y\,\ln\frac{\bar w}{z}\,+
 \bar Y\,\ln\frac{w }{\bar z}\,\Big),\label{VV3}
\end{eqnarray}
where the twisted chiral field $w$ and the chiral field $z$ are defined
by,
\begin{eqnarray}
 &&w=e^{-\bar \ID_+\ID_-u},\qquad \bar w=e^{+ \ID_+\bar \ID_-\bar u},\nonumber\\
 &&z=e^{i\,\bar \ID_+\bar \ID_-v},\qquad \bar z=e^{i\, \ID_+\ID_-\bar v}\,.
\end{eqnarray}
The T-dual model is obtained by integrating eq.~(\ref{VV3}) over the
gauge fields $X$, $Y$ and $\bar Y$. Doing this gives the equations of
motion,
\begin{eqnarray}
 X=i\,\ln\,\frac{w\,\bar l}{\bar w\,l},\qquad
 r=\frac{\bar w}{z},\qquad \bar r=\frac{w}{\bar z}.
\end{eqnarray}
Implementing this in eq.~(\ref{VV3}), we get the second order
potential,
\begin{eqnarray}
 V_3=\frac{k}{4\pi }\,\Big(\int^{\frac{z \bar z}{ w \bar w}} \frac{dq}{q}
\ln\left(1+q\right) - \frac{1}{2} \left( \ln w \bar w \right)^2+\cdots\,\Big),
\end{eqnarray}
where the omitted terms can be eliminated through a generalized K{\"a}hler
transformation. This expression indeed agrees with the one given in
eq.~(\ref{su2u1ctc4moduli}).

Let us now briefly discuss the inverse transformation where we will follow a slightly different 
strategy. Starting point is eq.~(\ref{su2u1ctc4moduli}). The $U(1)$ transformation acts
now on the fields as,
\begin{eqnarray}
\delta z= \varepsilon\, z,\qquad \delta w = \varepsilon\, w.
\end{eqnarray}
In order to make eq.~(\ref{su2u1ctc4moduli}) exactly invariant under this we add total derivative 
terms and our starting point becomes,
\begin{eqnarray} V(z,
w, \bar z, \bar w) &=& \frac{k}{4\pi }\,\Big(\int^{\frac{z \bar z}{ w \bar w}} \frac{dq}{q}
\ln\left(1+q\right) - \frac{1}{4} \left(\ln \frac{z}{\bar z}\right)^2
 + \frac{1}{4} \left(\ln \frac{w}{\bar w}\right)^2 \nonumber\\
&&  - \frac{1}{4} \left(\ln \frac{z\bar z}{w\bar w}\right)^2
+ \frac{ \alpha}{2} \ln \frac{z}{\bar z}\,\ln \frac{w}{\bar w}\Big)
,\label{su2u1ctc4modulimod}
\end{eqnarray}
where $\alpha\in\IR$. Any value of $\alpha$ is allowed as it multiplies a superspace
total derivative term. 
In order to gauge the isometry we introduce real gauge potentials $Y$, $\tilde Y$ and $\hat Y$,
transforming as,
\begin{eqnarray}
\delta Y=i\big( \varepsilon_z- \bar \varepsilon_{\bar z}\big),\qquad
\delta \tilde Y=i\big( \varepsilon_w- \bar \varepsilon_{\bar w}\big),\qquad
\delta \hat Y= \varepsilon_w+ \bar \varepsilon_{\bar w}
-\varepsilon_z- \bar \varepsilon_{\bar z} ,
\end{eqnarray}
where $\varepsilon_w$ is twisted chiral and $\varepsilon_z$ chiral.
The gauge invariant field strengths are given by $\bar\ID_+(Y-\tilde Y-i\hat Y)$, 
$\bar\ID_-(Y+\tilde Y-i\hat Y)$ and their complex conjugates. Using Lagrange multipliers we 
impose that the fieldstrengths vanish. Upon integrating by parts we obtain in this way the first order 
potential,
\begin{eqnarray} V&=& \frac{k}{4\pi }\,\Big( \frac{1}{4} \,Y^2
- \frac{1}{4}\,\tilde Y^2 
- \frac{ \alpha}{2}\, Y\tilde Y+
\int^{e^{\hat Y}} \frac{dq}{q}
\ln\left(1+q\right) - \frac{1}{4}\, \hat Y^2 \nonumber\\
&&+ \frac{i\,Y}{2}\,\ln \frac{lr}{\bar l \bar r}
+ \frac{i\,\tilde Y}{2}\ln \frac{\bar l r}{l\bar r}
+ \frac{\hat Y}{2}\ln l\bar l r\bar r\Big)
,\label{su2u1ctc4modulimod2}
\end{eqnarray}
where $l$ and $r$ form a semi-chiral multiplet. The equations of motion for $Y$, $\tilde Y$ and
$\hat Y$ give,
\begin{eqnarray}
Y&=& -i\,\ln\left(\frac{l}{\bar{l}}\right)^{\frac{1+\alpha}{1+\alpha^2}}  
\left(\frac{r}{\bar{r}}\right)^{\frac{1-\alpha}{1+\alpha^2}}
\,,\nonumber\\
\tilde Y&=&-i\,\ln\left(\frac{l}{\bar{l}}\right)^{\frac{1-\alpha}{1+\alpha^2}}  
\left(\frac{\bar r}{r}\right)^{\frac{1
+\alpha}{1+\alpha^2}}\,,\nonumber\\
e^{\hat Y}&=&\frac{1}{2X}\big(1-2X\pm\sqrt{1-4X}\big),\qquad\mbox{with }X=l\bar l r\bar r.
\end{eqnarray}
Implementing this in the first order potential gives,
\begin{eqnarray}
V(l,\bar l, r, \bar r)&=&\frac{k}{4\pi }\Big(
-\frac{(1+\alpha)^2}{2(1+\alpha^2)}\,\ln l\ln \bar{l}-
\frac{(1-\alpha)^2}{2(1+\alpha^2)}\,\ln r\ln \bar{r}+\nonumber\\
&&\frac{1-\alpha^2}{2(1+\alpha^2)}\,\big(\ln l\ln r+\ln \bar l\ln\bar r\big) -
\frac{3+\alpha^2}{2(1+\alpha^2)}\,\big(\ln l\ln\bar r+\ln \bar l\ln r\big) +\nonumber\\
&&\int^{l\bar l r\bar r}
\frac{dq}{q}\,\ln\big(1\pm\sqrt{1-4q}\big)\Big)\,,\label{piano8}
\end{eqnarray}
or alternatively,
\begin{eqnarray}
V(l,\bar l, r, \bar r)&=&\frac{k}{4\pi }\Big(
\frac{(1-\alpha)^2}{2(1+\alpha^2)}\,\ln l\ln \bar{l}+
\frac{(1+\alpha)^2}{2(1+\alpha^2)}\,\ln r\ln \bar{r}+\nonumber\\
&&\frac{3+\alpha^2}{2(1+\alpha^2)}\,\big(\ln l\ln r+\ln \bar l\ln\bar r\big) -
\frac{1-\alpha^2}{2(1+\alpha^2)}\,\big(\ln l\ln\bar r+\ln \bar l\ln r\big) -\nonumber\\
&&\int^{l\bar l r\bar r}
\frac{dq}{q}\,\ln\big(1\pm\sqrt{1-4q}\big)\Big)\,.\label{piano9}
\end{eqnarray}
One notices that eq.~(\ref{piano9}) is precisely the ``mirror transform'' of eq.~(\ref{piano8})
provided we send $\alpha$ to $-\alpha$ in eq.~(\ref{piano9}). Furthermore, putting $\alpha=-1$
in eq.~(\ref{piano8}) gives the potential in eq.~(\ref{piano}). Similarly we get eq.~(\ref{piano2}) when setting $\alpha=+1$ in eq.~(\ref{piano9}).
  
\subsection{Generalized K\"ahler geometry and related issues}
In order to keep the expressions transparent we will take throughout this section $\theta =\theta
'=\phi =\phi '=0$.
\subsubsection{The twisted chiral/chiral parametrization}
From the previous we know that the complex coordinates are related to the original coordinates on
the group by,
\begin{eqnarray}
w&=&\cos\psi\,e^{-\rho-i \varphi_1}, \nonumber\\
z&=&\sin\psi\,e^{-\rho+i \varphi_2}\,,
\end{eqnarray}
The generalized K{\"a}hler potential eq.~(\ref{su2u1ctc4moduli}), 
\begin{eqnarray} V_{w\neq 0}(
w, \bar w,z, \bar z) = \frac{k}{4\pi }\,\Big(\int^{\frac{z \bar z}{ w \bar w}} \frac{dq}{q}
\ln\left(1+q\right) - \frac{1}{2} \left( \ln w \bar w \right)^2 \Big)\,,
\label{su2u1ctc4modulibb}
\end{eqnarray}
is everywhere well defined except when $w=0$ (or $\psi =\pi /2$). However
-- as noted before in \cite{Sevrin:2008tp} and \cite{Hull:2008vw} -- we can consider the 
``mirror'' potential (see eq.~(\ref{mirrorpot})),
\begin{eqnarray} V_{z\neq 0}(w,
\bar w, z,\bar z) = \frac{k}{4\pi }\,\Big(-\int^{\frac{w \bar w}{ z \bar z}} \frac{dq}{q}
\ln\left(1+q\right) + \frac{1}{2} \left( \ln z \bar z \right)^2 \Big)\,,\label{WZWpot2b}
\end{eqnarray}
which is now well defined as long as $z\neq 0$ (or $\psi\neq 0$). The potentials in the two patches are 
related by a generalized K\"ahler transformation,
\begin{eqnarray}
V_{w\neq 0}-V_{z\neq 0}=-\frac{k}{4\pi }\,\ln\big( z\bar z\big)\ln\big( w\bar w\big)\,.
\end{eqnarray}

We now turn to the pure spinors and start from the potential in eq.~(\ref{su2u1ctc4modulibb}) 
-- {\em i.e.} we work in the patch where $w\neq 0$ (or $\psi\neq \pi/2)$. The two pure spinors $\phi_+$ 
and $\phi_-$ are respectively given by
\begin{eqnarray}
\phi_+ = d\bar z \wedge e^{i\, \Omega^+ + \Xi^+}=
d\bar z \wedge e^{-b\wedge}\, e^{\Lambda^+ }, \nonumber\\
\phi_- = d\bar w \wedge e^{i\, \Omega^- + \Xi^-}=
d\bar w \wedge e^{-b\wedge}\, e^{\Lambda^-},\label{psctc}
\end{eqnarray}
where,
\begin{eqnarray}
i\, \Omega^+ + \Xi^+ =  \frac{k}{ 8 \pi( w \bar w + z \bar z)} \left( - 2 dw \wedge d \bar w - 
\frac{\bar z}{ w}  dw \wedge dz + \frac{z}{w} dw \wedge d \bar z \right. \nonumber \\
 \left. + \frac{\bar z}{\bar w} d \bar w \wedge dz +3 \frac{z}{\bar w} d \bar w \wedge d \bar z - 2 dz 
\wedge  d \bar z  \right),
\end{eqnarray}
\begin{eqnarray}
i\, \Omega^- + \Xi^- = \frac{k}{ 8 \pi( w \bar w + z \bar z)} \left(  2 dw \wedge d \bar w - 
\frac{\bar z}{ w}  dw \wedge dz + \frac{z}{w} dw \wedge d \bar z \right. \nonumber \\
 \left. -3 \frac{\bar z}{\bar w} d \bar w \wedge dz - \frac{z}{\bar w} d \bar w \wedge d \bar z - 2 dz 
\wedge  d \bar z  \right).
\end{eqnarray}
We can also undo the $b$-transform and obtain in this way the $H$-closed pure spinors $\phi_+ = d\bar z \wedge  e^{\Lambda_+}$, $\phi_- = d\bar w \wedge  e^{\Lambda_-}$, where,
\begin{eqnarray}
\Lambda_+ &=& -\frac{k}{4\pi} \frac{1}{w \bar w + z \bar z} \left( dw \wedge d\bar w + d z \wedge d\bar z  - 2 \frac{z}{\bar w}d \bar w \wedge d\bar z  \right), \label{Lampsimctcp1}\\
\Lambda_- &=& \frac{k}{4\pi} \frac{1}{w \bar w + z \bar z} \left( dw \wedge d\bar w - d z \wedge d\bar z  - 2 \frac{\bar z}{\bar w}d \bar w \wedge d z  \right), \label{Lammsimctcp1}
\end{eqnarray}
and,
\begin{eqnarray}
b = \frac{k}{8 \pi} \frac{1}{z \bar z + w \bar w}\left(  \frac{\bar z}{w} dw \wedge dz - \frac{z}{w} dw \wedge d \bar z - \frac{ z}{\bar w} d \bar w \wedge d \bar z + \frac{z}{\bar w} d \bar w \wedge d \bar z  \right).
\end{eqnarray}
Both $\phi_+$ and $\phi_-$ are well defined as long as $w\neq 0$. The generalized K\"ahler structure is 
of type $(1,1)$ and no type changing occurs (as was
expected). For the Mukai pairings, eq.~(\ref{mukaip}), we find,
\begin{eqnarray}
\big( \phi_+, \bar\phi_+\big)=\big( \phi_-, \bar \phi_-\big)=- \frac{k}{2 \pi}\, \frac{1}{z \bar{z}
+w \bar{w}}\,=- \frac{k}{2 \pi}\, e^{2 \rho}\,,
\end{eqnarray}
which vanishes nowhere and which  satisfies the generalized Calabi-Yau condition 
eq.~(\ref{CYcd1}) as well.  

In the other patch -- where $z\neq 0$ or $\psi\neq 0$ -- described by the potential in 
eq.~(\ref{WZWpot2b}), we find again pure spinors of the form given in eq.~(\ref{psctc}) but now with,
\begin{eqnarray}
i\, \Omega^+ + \Xi^+ =  \frac{k}{ 8 \pi( w \bar w + z \bar z)} \left( - 2 dw \wedge d \bar w + \frac{\bar 
w}{ z}  dw \wedge dz - \frac{\bar w}{\bar z} dw \wedge d \bar z \right. \nonumber \\
 \left. - \frac{w}{z} d \bar w \wedge dz -3 \frac{w}{\bar z} d \bar w \wedge d \bar z - 2 dz \wedge  d 
\bar z  \right), \nonumber\\
i\, \Omega^- + \Xi^- =  \frac{k}{ 8 \pi( w \bar w + z \bar z)} \left(  2 dw \wedge d \bar w + \frac{\bar 
w}{ z}  dw \wedge dz - \frac{\bar w}{\bar z} dw \wedge d \bar z \right. \nonumber \\
 \left. + 3 \frac{w}{z} d \bar w \wedge dz + \frac{w}{\bar z} d \bar w \wedge d \bar z - 2 dz \wedge  d 
\bar z  \right),
\end{eqnarray}
and undoing the $b$-transform we obtain the $H$-closed pure spinors $\phi_+ = d\bar z \wedge  e^{\Lambda_+}$, $\phi_- = d\bar w \wedge  e^{\Lambda_-}$, with,
\begin{eqnarray}
\Lambda_+ &=& -\frac{k}{4\pi} \frac{1}{w \bar w + z \bar z} \left( dw \wedge d\bar w + d z \wedge d\bar z  + 2 \frac{w}{\bar z}d \bar w \wedge d\bar z  \right), \label{Lampsimctcp2}\\
\Lambda_- &=& \frac{k}{4\pi} \frac{1}{w \bar w + z \bar z} \left( dw \wedge d\bar w - d z \wedge d\bar z  + 2 \frac{w}{z}d \bar w \wedge d z  \right).\label{Lammsimctcp2}
\end{eqnarray} 
Again this is well defined as long as $z\neq 0$ and it satisfies the generalized Calabi-Yau condition.
However, when comparing the expressions for the pure spinors in both patches one finds that they are
not globally well defined thus leading to the conclusion that this model, while $N=(2,2)$ superconformally
invariant does not provide for a consistent supergravity background. 

\subsubsection{The semi-chiral parametrization}\label{section:432}
We now turn to the semi-chiral parameterization. As no chiral or twisted chiral fields
are involved, the two-form $\Omega$, eq.~(\ref{globa1}) can be introduced. At first sight
it looks globally well defined. However, if it were so then by virtue of eq.~(\ref{globa2}) the 
Kalb-Ramond two-form $b$ would be globally defined as well, implying that the torsion 3-form 
is exact. This is clearly not true on a group manifold! This is resolved once one realizes 
that type changing occurs. Indeed as we shall see we get that there are two $T^2$ submanifolds
where $[J_+,J_-]=0$, so $[J_+,J_-]$ is not invertible and as a consequence $ \Omega$ is 
not defined in these loci. This situation has to be contrasted with 
the previous case. There both chiral {\em and} twisted chiral coordinates are present and 
none of the expressions for $\Omega^+$ (eq.~(\ref{opglob})) , $\Omega^-$ 
(eq.~(\ref{omglob}))  or $ \Omega$ (eq.~(\ref{globa1}))  are well defined.

%: glob

In order to proceed we start from the potential in eq.~(\ref{Vsemi}) but make a coordinate 
transformation by taking the logarithm of the original coordinates. With this we get,
\begin{eqnarray}
 V_{\psi\neq 0}(l,\bar l,r,\bar r)=\frac{k}{4\pi }\Big(\big(l-\bar r\big)\big(\bar l -r\big)-\int^{r+\bar r}dq\,
 \ln\big(1+e^q\big)\Big),\label{Vsemilog}
\end{eqnarray}
where the coordinates are explicitly given by,
\begin{eqnarray}
l=\ln\left(\cos \psi\,e^{-\rho-i\,\varphi_1}\right),\qquad r=\ln\left(\cot \psi\,e^{i \,\varphi_1-i\, \varphi_2}\right).
\label{cos1}
\end{eqnarray}  
The potential is well defined as long as $\psi\neq 0$. The complex structures are,
\begin{eqnarray}
 J_+=\left(\begin{array}{cccc}
   +i & 0 & 0 & 0 \\
   0 &-i & 0 & 0 \\
   0 & -2i & +i&0\\
   +2i&0&0&-i
 \end{array}   \right),\qquad
J_-=\left(\begin{array}{cccc}
   i&0&0&-\frac{2i}{1+e^{r+\bar r}} \\
   0& -i& \frac{2i}{1+e^{r+\bar r}}&0\\
   0 & 0 & +i & 0 \\
   0 & 0 & 0 & -i
 \end{array}   \right).\label{thecomstrlog}
\end{eqnarray}
The metric is given by,
\begin{eqnarray}
ds^2= \frac{k}{2 \pi}\left(dl\,d \bar{l}+ \frac{1}{1+e^{r+ \bar{r}}}\left(dr\,d \bar{r}
-dl\,dr-d \bar{l}\,d \bar{r}
\right)\right).
\end{eqnarray}
and the torsion 3-form by,
\begin{eqnarray}
H=\frac{k}{4\pi }\Big(\frac{e^{r+\bar r}}{(1+e^{r+\bar r})^2}\,dl\wedge dr\wedge d\bar r-
\frac{e^{r+\bar r}}{(1+e^{r+\bar r})^2}\,  d\bar l\wedge dr\wedge d\bar r \Big).
\end{eqnarray}
One also computes,
\begin{eqnarray}
\det \big(J_++J_-\big)&=& \frac{16\,e^{2( r+ \bar{r})}}{(1+e^{r +\bar{r}})^2}=
16\cos^4\psi, \nonumber\\
\det \big(J_+-J_-\big)&=&\frac{16}{(1+e^{r +\bar{r}})^2}=
16\sin^4\psi,
\end{eqnarray}
so we anticipate type changing to occur at $\psi=\pi/2$ where the locus will be parameterized by a twisted chiral field.

The second coordinate patch (well defined as long as $ \psi\neq \pi/2$) is described by the potential,
\begin{eqnarray}
 V_{\psi\neq \frac \pi 2}(l,\bar l,r,\bar r)=\frac{k}{4\pi }\Big(-\big(l- r\big)\big(\bar l -\bar r\big)+\int^{r+\bar r}dq\,
 \ln\big(1+e^q\big)\Big),\label{Vsemilog2}
\end{eqnarray}
which is the ``mirror'' of  
eq.~(\ref{Vsemilog}) which as we saw can be obtained
by Legendre transforming eq.~(\ref{Vsemilog}) w.r.t. $l$ and $\bar{l}$. In this case the coordinates are 
given by,
\begin{eqnarray}
l=\ln\left(\sin \psi\,e^{-\rho+i\,\varphi_2}\right),\qquad r=\ln\left(\tan \psi\,e^{-i \,\varphi_1+i\, 
\varphi_2}\right).
\label{cos2}
\end{eqnarray}  
As long as the superspace description only involves chiral and twisted chiral coordinates
one has that on the overlap of two coordinate patches the two expressions for the K\"ahler 
potential are related through a generalized K\"ahler transformation. The present example 
clearly shows that this changes once semi-chiral coordinates are present as well. In this 
case one might need a Legendre transform when relating the potentials on the 
overlap.
The complex structures are now given by,
\begin{eqnarray}
 J_+=\left(\begin{array}{cccc}
   +i & 0 & 0 & 0 \\
   0 &-i & 0 & 0 \\
   2i &0 & -i&0\\
   0&-2i&0&i
 \end{array}   \right),\qquad
J_-=\left(\begin{array}{cccc}
  - i&0&\frac{2i}{1+e^{r+\bar r}}&0 \\
   0& i&0&- \frac{2i}{1+e^{r+\bar r}}\\
   0 & 0 & +i & 0 \\
   0 & 0 & 0 & -i
 \end{array}   \right).\label{thecomstrlog2}
\end{eqnarray}
The metric and torsion become,
\begin{eqnarray}
ds^2&=& \frac{k}{2 \pi}\left(dl\,d \bar{l}+ \frac{1}{1+e^{r+ \bar{r}}}\left(dr\,d \bar{r}
-dl\,d\bar{r}-d \bar{l}\,d r
\right)\right),\nonumber\\
H&=&-\frac{k}{4\pi }\Big(\frac{e^{r+\bar r}}{(1+e^{r+\bar r})^2}\,dl\wedge dr\wedge d\bar r-
\frac{e^{r+\bar r}}{(1+e^{r+\bar r})^2}\,  d\bar l\wedge dr\wedge d\bar r \Big).
\end{eqnarray}
Using the complex structures one finds,
\begin{eqnarray}
\det \big(J_++J_-\big)&=& \frac{16}{(1+e^{r +\bar{r}})^2}=
16\cos^4\psi, \nonumber\\
\det \big(J_+-J_-\big)&=&\frac{16\,e^{2( r+ \bar{r})}}{(1+e^{r +\bar{r}})^2}=
16\sin^4\psi,
\end{eqnarray}
and the type will change at $\psi=0$ where we will have a description in terms of a chiral field.

For now we work on the first patch where $\psi\neq 0$ described by the potential in 
eq.~(\ref{Vsemilog}). The pure spinors are readily calculated and give,
\begin{eqnarray}
\phi_+ =e^{i\, \Omega^+ + \Xi^+}, \nonumber\\
\phi_- = e^{i\, \Omega^- + \Xi^-},\label{pspss3}
\end{eqnarray}
with $ i\, \Omega^+ + \Xi^+ $, 
\begin{eqnarray}
 i\, \Omega^+ + \Xi^+  = \frac{k}{8 \pi} \left( 2 dl\wedge d \bar l - dl \wedge dr + 3 d \bar l \wedge 
d\bar r - \frac{2}{1+ e^{r+ \bar r}} dr \wedge d\bar r  \right),
\end{eqnarray}
 and $ i\, \Omega^- + \Xi^-$,
 \begin{eqnarray}
i\,  \Omega^- + \Xi^-  = \frac{k}{8 \pi} \left( - 2 dl\wedge d \bar l -  dl \wedge dr - d \bar l \wedge 
d\bar r - \frac{2}{1+ e^{r+ \bar r}} dr \wedge d\bar r  \right).
 \end{eqnarray}
Using these expressions, one computes the Mukai pairings,
\begin{eqnarray}
\big( \phi_+,\bar \phi_+\big)&=& \frac{k^2}{4 \pi^2}\, \frac{e^{r+\bar r}}{1+e^{r+\bar r}}\,=
\frac{k^2}{4 \pi^2}\,\cos^2\psi\,, \nonumber\\
\big( \phi_-,\bar \phi_-\big)&=& \frac{k^2}{4 \pi^2}\, \frac{1}{1+e^{r+\bar r}}\,=
\frac{k^2}{4 \pi^2}\,\sin^2\psi\,.\label{Muksc}
\end{eqnarray}
We notice two things here. Firstly the generalized Calabi-Yau condition eq.~(\ref{CYcd1}) is not satisfied 
here while the weaker condition eq.~(\ref{CYcd2}) is satisfied. Secondly, we work here on the patch 
where $\psi\neq 0$. When $\psi=\pi/2$ (where type-changing occurs) one gets that $( \phi_+,\bar 
\phi_+)=0$! 

In order to investigate this we introduce new coordinates $w$ and $z$,
\begin{eqnarray}
w=e^l,\qquad z=e^{-r+\bar l}.
\end{eqnarray}
In these coordinates one gets that the complex structure $J_+$ is diagonalized with eigenvalues $+i$ on 
$dw$ and $dz$ while $J_-$ is now given by,
\begin{eqnarray}
J_-= \frac{1}{w\bar w+ z\bar z} \left(\begin{array}{cccc}
   +i(w\bar w- z\bar z )& 0 & 0 & +2iwz \\
   0 &-i(w\bar w- z\bar z) & -2i\bar w\bar z & 0 \\
   0 & -2i w z& +i(w\bar w- z\bar z)&0\\
   +2i\bar w\bar z&0&0&-i(w\bar w- z\bar z)
 \end{array}   \right),
\end{eqnarray}
where rows and columns are labeled in the order $w$, $\bar w$, $z$ and $\bar z$.
The metric $g$ in these coordinates is given by eq.~(\ref{gchitwi}) and the Kalb-Ramond 
form $b$ -- in a gauge given by eq.~(\ref{bgauge}) -- is,
\begin{eqnarray}
b= -\frac{k}{8\pi}\, \frac{w\bar w- z\bar z}{w\bar w+ z\bar z}
\left(\frac{dw\wedge dz}{wz}+\frac{d\bar w\wedge d\bar z}{\bar w\bar z}\right)\,,
\end{eqnarray} 
which results in a torsion 3-form $H=db$ as in eq.~(\ref{Hchtwch}). 
Performing the coordinate transformation on the pure spinors eq.~(\ref{pspss3}) results in,
\begin{eqnarray}
\phi_\pm= \sqrt{w\bar w z \bar z}\,e^{-b\wedge }\,e^{\Lambda_\pm},\label{psps1}
\end{eqnarray}
where,
\begin{eqnarray}
\Lambda_+&=&- \frac{k}{4\pi}\, \frac{1}{w\bar w+ z\bar z}
\left(dw\wedge d\bar w+dz\wedge d\bar z+2 \frac{w}{\bar z}\,d\bar w\wedge d\bar z\right)\,, 
\nonumber\\
\Lambda_-&=&- \frac{k}{4\pi}\, \frac{1}{w\bar w+ z\bar z}
\left(dw\wedge d\bar w+dz\wedge d\bar z-2 \frac{z}{\bar w}\,d\bar w\wedge d\bar z\right)\,,
\end{eqnarray}
and one verifies that,
\begin{eqnarray}
de^{ \Lambda_\pm}=H\wedge e^{ \Lambda_\pm}.
\end{eqnarray}
The multiplicative prefactor $\sqrt{w\bar w z \bar z}$ in eq.~(\ref{psps1}) arises from the 
fact that pure spinors transform as a density under coordinate transformations and is indeed needed 
to keep the Mukai pairing invariant under a coordinate transformation. The $b$-transform in  
eq.~(\ref{psps1}) has obviously no influence on the Mukai pairing. 

In these coordinates, the type changing locus $\psi=\pi/2$ corresponds to $w=0$ and we clearly see from 
the previous that then $J_-=-J_+$ and ${\cal J}_+$ and ${\cal J_-}$ respectively become of the symplectic 
and complex type respectively. This becomes manifest if we rescale the pure spinors in an appropriate 
way\footnote{In order to simplify the expressions we performed a $b$-transform as well. Put differently, we once more consider a generalized K\"ahler geometry where the integrability are determined using the $H$-twisted Courant bracket.},
\begin{eqnarray}
\phi_+&=&e^{\Lambda_+}, \nonumber\\
\phi_-&=&\bar w\, e^{\Lambda_-}.
\end{eqnarray}
This shows that as long as $w\neq 0$ the generalized K\"ahler geometry is of type $(0,0)$. When 
$w \rightarrow 0$ the type jumps to $(0,2)$, and the pure spinors reduce to their canonical forms 
eq.~(\ref{canpusp}) ($\phi_+$ remains symplectic but $\phi_-$ becomes complex). Note that while $\phi_+$ is still $H$-closed, this is not true for $\phi_-$ anymore.
The Mukai pairings are now,
\begin{eqnarray}
(\phi_+,\bar \phi_+)&=& \frac{k^2}{4 \pi^2}\,\frac{1}{w\bar w+ z\bar z}\, \frac{1}{z\bar z} 
\nonumber\\
(\phi_-,\bar \phi_-)&=& \frac{k^2}{4 \pi^2}\,\frac{1}{w\bar w+ z\bar z}\,, 
\end{eqnarray}
which is now non-degenerate even if $w=0$. From this analysis of a superconformally invariant model it is 
clear that once type changing is involved a proper definition of the pure spinors becomes more intricate.

We now turn to the other coordinate patch $\psi\neq\pi/2$ where the potential is given by 
eq.~(\ref{Vsemilog2}) and the coordinates by eq.~(\ref{cos2}). The pure spinors are still given by eq.~(\ref{pspss3}) but the 
relevant 2-forms are now,
\begin{eqnarray}
 i\, \Omega^+ + \Xi^+  = \frac{k}{8 \pi} \left( -2 dl\wedge d \bar l - dl \wedge d\bar r - d \bar l \wedge 
d r + \frac{2}{1+ e^{r+ \bar r}} dr \wedge d\bar r  \right),
\end{eqnarray}
 and,  
 \begin{eqnarray}
i\,  \Omega^- + \Xi^-  = \frac{k}{8 \pi} \left(  2 dl\wedge d \bar l -  dl \wedge d\bar r + 3d \bar l \wedge 
d r + \frac{2}{1+ e^{r+ \bar r}} dr \wedge d\bar r  \right).
 \end{eqnarray}
Calculating the Mukai pairings results in,
\begin{eqnarray}
\big( \phi_+,\bar \phi_+\big)&=&- \frac{k^2}{4 \pi^2}\, \frac{1}{1+e^{r+\bar r}}\,=-
\frac{k^2}{4 \pi^2}\,\cos^2\psi\,, \nonumber\\
\big( \phi_-,\bar \phi_-\big)&=&- \frac{k^2}{4 \pi^2}\, \frac{e^{r+\bar r}}{1+e^{r+\bar r}}\,=-
\frac{k^2}{4 \pi^2}\,\sin^2\psi\,.\label{Muksc2}
\end{eqnarray}
Once again we notice that the generalized Calabi-Yau condition eq.~(\ref{CYcd1}) is not satisfied while
the model is UV-finite eq.~(\ref{CYcd2}) (and as it describes a WZW-model it is $N=(2,2)$ 
superconformally invariant as well) and the Mukai pairing is degenerate at the type changing
locus $\psi=0$.  Making a coordinate transformation,
\begin{eqnarray}
w=e^l,\qquad z=e^{\bar l-\bar r},
\end{eqnarray}
allows one to analyze type changing in a similar way as above. The complex structure $J_+$ is now diagonalized while $J_-$ is given by,
\begin{eqnarray}
J_-= \frac{1}{w\bar w+ z\bar z} \left(\begin{array}{cccc}
   -i(w\bar w- z\bar z )& 0 & 0 & -2iwz \\
   0 &+ i(w\bar w- z\bar z) & 2i\bar w\bar z & 0 \\
   0 & 2i w z& - i(w\bar w- z\bar z)&0\\
   2i\bar w\bar z&0&0&+ i(w\bar w- z\bar z)
 \end{array}   \right).
\end{eqnarray}
When $w=0$ (or $\psi=0$) $J_+=J_-$ and the type changes.
One finds that the role of $\phi_+$ and 
$\phi_-$ get interchanged\footnote{This should not surprise us, as this interchanging is a manifestation of the mirror symmetry between the two patches.},
\begin{eqnarray}
\phi_+&=&\bar w\, e^{\Lambda_-}, \nonumber\\
\phi_-&=&e^{\Lambda_+}.
\end{eqnarray}
Again it is clear that at $w=0$ ($\psi=0$) the type jumps from $(0,0)$ to $(2,0)$ and $\phi_+$ becomes of complex type while $\phi_-$ remains symplectic. The Mukai pairings are now non-degenerate, 
but $\phi_+$ is not $H$-closed anymore nor is the generalized Calabi-Yau condition satisfied.

\subsection{Isometries}
We introduce,
\begin{eqnarray}
h_-=e^{i\epsilon_-\sigma_0+i\eta_-\sigma_3},\qquad h_+=e^{i\epsilon_+\sigma_0+i\eta_+\sigma_3},
\label{S3S1aff}
\end{eqnarray}
and consider the transformation $g\rightarrow h_-gh_+$ with $g$ given in eq.~(\ref{su2u1coor}). Its effect on the chiral and twisted chiral coordinates is given by,
\begin{eqnarray}
w\rightarrow e^{-\phi_--\phi_+}\,w,\qquad z\rightarrow e^{-\bar\phi_--\phi_+}\,z,
\end{eqnarray}
where $\phi_\pm=\epsilon_\pm+i\eta_\pm$. Compatibility of these transformation rules with the superfield constraints requires that,
\begin{eqnarray}
&&\ID_+\phi_-=\bar\ID_+\phi_-=\ID_-\phi_-=0,\nonumber\\
&&\ID_-\phi_+=\bar\ID_-\phi_+=\bar\ID_+\phi_+=0,
\end{eqnarray}
which imply that $\partial_\pp\, \phi_-=\partial_=\,\phi_+=0$. This clearly shows that these transformations correspond to (abelian) affine symmetries. The potential eq.~(\ref{su2u1ctc4moduli}) is indeed -- modulo a superspace total derivative term -- invariant under these affine symmetries.

The effect of the transformations eq.~(\ref{S3S1aff}) on the semi-chiral coordinates $l$ and $r$ is given by,
\begin{eqnarray}
l\rightarrow e^{-\phi_--\phi_+}\,l,\qquad r\rightarrow e^{\phi_+-\bar\phi_+}\,r.
\end{eqnarray} 
Compatibility of the symmetry transformations with the superfield constraints requires,
\begin{eqnarray}
\bar\ID_+\phi_-=\bar\ID_+\phi_+=0,\qquad 
\bar\ID_-\phi_+=\bar\ID_-\bar\phi_+,\qquad
\ID_-\phi_+=\ID_-\bar\phi_+.
\end{eqnarray} 
This is not sufficient to ensure the invariance (modulo total derivative terms) of the potential eq.~(\ref{Vsemi}). In order to achieve this we need to require,
\begin{eqnarray}
&&\ID_+\phi_-=\bar\ID_+\phi_-=\bar\ID_-\phi_-=0,\nonumber\\
&&\ID_-\phi_+=\bar\ID_-\phi_+=\bar\ID_+\phi_+=0,
\end{eqnarray}
which once more implies  $\partial_\pp\, \phi_-=\partial_=\,\phi_+=0$.

Let us now briefly reconsider the isometries which do not belong to the maximal torus.
As discussed in section 3 these do not leave the complex structures invariant and as a 
consequence the transformed fields will not be compatible with the original superfield
constraints. For concreteness we will consider an infinitesimal transformation $g \rightarrow h_-\,g$ and $h_-={\bf 1}+ \delta h_-$ where,
\begin{eqnarray}
\delta h_-=\left(
\begin{array}{cc}
0& \varepsilon\\
-\bar \varepsilon&0\end{array}
\right)\,.
\end{eqnarray}
The variation of the group coordinate is $\delta \rho=0$ and,
\begin{eqnarray}
\delta \psi &=&\frac 1 2 \, \varepsilon\, e^{-i( \varphi_1+ \varphi_2)}+
\frac 1 2 \, \bar\varepsilon\, e^{i( \varphi_1+ \varphi_2)}\,, \nonumber\\
\delta \varphi_1&=&\frac i 2 \tan \psi \big( \varepsilon\,e^{-i( \varphi_1+ \varphi_2)}
-\bar \varepsilon\,e^{i( \varphi_1+ \varphi_2)}\big)\,, \nonumber\\
\delta \varphi_2&=&-\frac i 2 \cot \psi \big( \varepsilon\,e^{-i( \varphi_1+ \varphi_2)}
-\bar \varepsilon\,e^{i( \varphi_1+ \varphi_2)}\big)\,,
\end{eqnarray}
which using eq.~(\ref{aux1}) results in,
\begin{eqnarray}
\delta w=-\bar \varepsilon \,z,\qquad \delta z = \varepsilon\,w.\label{odtrans}
\end{eqnarray} 
The above transformation leaves $J_+$ invariant and as a consequence 
eq.~(\ref{odtrans}) respects the constraints following from $J_+$, {\em i.e.}
the equations are annihilated by $\bar \ID_+$. However as $J_-$ is not invariant, the transformation rules are incompatible with the constraints following from it. Acting with $\hat D_-$ on the transformed coordinates, $w'=w+ \bar \varepsilon\,z$ and $z'=z-\varepsilon\,w$, one finds that the complex structure is deformed to,
\begin{eqnarray}
J_- =\left(
\begin{array}{cccc}
-i&0&-2i\,\bar \varepsilon&0\\
0&i&0&2i \,\varepsilon\\
-2i \,\varepsilon&0&i&0\\
0&2i\,\bar \varepsilon&0&-i\end{array}
\right)\,,\label{csdeff}
\end{eqnarray}
where rows and columns are labeled in the order $w'$, $\bar w'$, $z'$, $\bar z'$. This is
perfectly consistent with eq.~(\ref{csiso}). Eq.~(\ref{csdeff}) is preserved as long as $\bar\ID_+ \varepsilon=\ID_+ \varepsilon=\ID_- \varepsilon=0$ implying that $ \partial _\pp \,\varepsilon=0$. 

\section{$SU(2)\times SU(2)$}\label{sect:5}
%THIS SECTION STILL HEAVILY UNDER CONSTRUCTION!
\subsection{The complex geometry of the group}\label{sect:51}
Another interesting example is the WZW-model on $SU(2)\times SU(2)$, which was already briefly considered in \cite{Sevrin:1996jr}. We parameterize a group element $g$ as,
\begin{eqnarray}
g= \left(
\begin{array}{cccc} \cos \psi_1 \, e^{i\varphi _{11}} & \sin \psi_1 \,e^{i\varphi _{12}}&0&0\\
-\sin\psi_1\, e^{-i\varphi _{12}} & \cos\psi_1\, e^{-i\varphi _{11}}&0&0\\
 0&0&\cos \psi_2 \, e^{i\varphi _{21}} & \sin \psi_2 \,e^{i\varphi _{22}}\\
0&0&-\sin\psi_2\, e^{-i\varphi _{22}} & \cos\psi_2\, e^{-i\varphi _{21}}
\end{array}\right),
\label{su2su2coor}
\end{eqnarray}
where $\varphi _{11},\, \varphi_{12},\, \varphi _{21},\, \varphi_{22} \in \IR\,\mbox{mod}\,2 \pi $ and
$\psi_1,\psi_2\in[0, \pi /2]$. The local topology of the group manifold $SU(2)\times SU(2)$ thus corresponds to $S^3\times S^3$, where the first three-sphere is parametrized by $\varphi_{11}$, $\varphi_{12}$ and $\psi_1$, and the second three-sphere by $\varphi_{21}$, $\varphi_{22}$ and $\psi_2$. In these coordinates the metric on $S^3\times S^3$ is given by
\begin{eqnarray}
ds^2 = \frac{k}{2 \pi} \left(  d \psi_1^2 + \cos^2 \psi_1 \, d\varphi_{11}^2 + \sin^2 \psi_1 \, d \varphi_{12}^2+d \psi_2^2 + \cos^2 \psi_2 \, d\varphi_{21}^2 + \sin^2 \psi_2 \, d \varphi_{22}^2 \right), \label{S3S3met}
\end{eqnarray}
and the torsion three-form reads
\begin{eqnarray}
H = \frac{k}{2 \pi} \left( \sin 2 \psi_1 \, d\varphi_{11} \wedge d\varphi_{12} \wedge d \psi_1 + \sin 2 \psi_2 \, d\varphi_{21} \wedge d\varphi_{22} \wedge d \psi_2  \right). \label{S3S3tor}
\end{eqnarray}
At the endpoints $\psi_1=0$ ($\psi_2=0$) or $\psi_1=\pi/2$ ($\psi_2=\pi/2$) the group manifold is pinched down to $S^1\times S^3$ ($S^3\times S^1$), as can be clearly seen from the metric (\ref{S3S3met}) and the torsion three-form (\ref{S3S3tor}).
 
We write the Lie algebra generators of $SU(2)\times SU(2)$ as,
\begin{eqnarray}
h =\frac 1 2 \,  \left(\begin{array}{cccc} 1&0&0&0\\
0&-1&0&0\\0&0&i&0\\0&0&0&-i
\end{array} \right), \quad
e_1 = \left(\begin{array}{cccc} 0&1&0&0\\
0&0&0&0\\0&0&0&0\\0&0&0&0
\end{array} \right),\quad
e_2 = \left(\begin{array}{cccc} 0&0&0&0\\
0&0&0&0\\0&0&0&1\\0&0&0&0
\end{array} \right),
\end{eqnarray}
and $\bar h=h^\dagger$, $\bar e_1=e_1^\dagger$ and  $\bar e_2=e_2^\dagger$. 

Modulo group conjugation we can make two choices for the complex structure on the Lie algebra:
$\IJ_1=\mbox{diag} \big(+i,+i,+i,-i,-i,-i\big)$ or $\IJ_2=\mbox{diag} \big(-i,+i,+i,+i,-i,-i\big)$ where we labeled rows and columns in the order $h,\,e_1,\,e_2,\,\bar h,\,\bar e_1,\,\bar e_2$. So we get the following options for the complex structures on the group,
\begin{enumerate}
  \item $J_+^a{}_b=L^a_C\,\IJ_1^C{}_D\,L^D_b$,\quad
  $J_-^a{}_b=R^a_C\,\IJ_1^C{}_D\,R^D_b\,$.
   \item $J_+^a{}_b=L^a_C\,\IJ_1^C{}_D\,L^D_b$,\quad
  $J_-^a{}_b=R^a_C\,\IJ_2^C{}_D\,R^D_b\,$.
\end{enumerate}
For the first choice one finds that the eigenvalues of $J_++J_-$ are given by
$\pm 2 i$ and $\pm 2 i \cos \psi_1\cos \psi_2$ where the last two eigenvalues are two-
folded degenerated. The eigenvalues of $J_+-J_-$ are $0$ and 
$\pm i \sqrt{3-2\cos 2 \psi_1\cos^2\psi_2-\cos 2 \psi_2}$ where each eigenvalue is two-
folded degenerated. So for this choice for the complex structures the superspace 
description will be given in terms of a chiral superfield and a semi-chiral multiplet. Looking a bit closer at the eigenvalues, we expect that type changing occurs for three different loci: $\psi_1=\pi/2$, $\psi_2=\pi/2$ and $\psi_1=\psi_2=0$. 

If we take the second choice for the complex structures we get for the eigenvalues of 
$J_++J_-$ $0$ and 
$\pm i \sqrt{3+2\cos 2 \psi_1\cos^2\psi_2+\cos 2 \psi_2}$ (each two-
folded degenerated). The eigenvalues of $J_+-J_-$ are 
$\pm 2 i$ and $\pm 2 i \sin \psi_1\sin \psi_2$ (the last two eigenvalues two-folded degenerated). So here we will need one twisted chiral superfield and one semi-chiral multiplet. Looking a bit closer again at the eigenvalues, type changing will occur for three different loci: $\psi_1=0$, $\psi_2=0$ and $\psi_1=\psi_2=\pi/2$. 

%: s3s3

We now turn to the first choice for the complex structurs which results in the potential,
\begin{eqnarray}
 V(l,\bar l,r,\bar r,z,\bar z)&=&\frac{k}{4\pi }\Big(-(l-r)( \bar{l}- \bar{r})+\int^{r+\bar r}
dq\,
 \ln\big(1+e^q\big)+ \nonumber\\
&&\qquad \int^{z+ \bar{z}+i(l-\bar l)}dq\,
 \ln\big(1+e^q\big)\Big),\label{s3s3V}
\end{eqnarray}
where,
\begin{eqnarray}
&&l=\ln\sin \psi_2+i\,\ln\cos \psi_1 + \varphi_{11}+i\, \varphi_{22}\,,  \nonumber\\
&&r=\ln\tan\psi_2+i(\varphi_{22}-\varphi_{21})\,,\nonumber\\
&&z=\ln\sin \psi_1-i\ln\sin \psi_2+\varphi_{22}+i \,\varphi_{12}\,.
\end{eqnarray}
This potential is well defined for $ \psi_1,\,\psi_2\neq\pi/2$.

For the second choice of the complex structures we find the ``mirror'' potential,
\begin{eqnarray}
 V(l,\bar l,r,\bar r,w,\bar w)&=&\frac{k}{4\pi }\Big((l-\bar r)( \bar{l}- r)-\int^{r+\bar r}
dq\,
 \ln\big(1+e^q\big)- \nonumber\\
&&\qquad \int^{w+ \bar{w}+i(l-\bar l)}dq\,
 \ln\big(1+e^q\big)\Big),\label{s3s3V2}
\end{eqnarray}
where,
\begin{eqnarray}
&&l=\ln\cos \psi_2+i\,\ln\sin \psi_1 - \varphi_{12}-i\, \varphi_{21}\,,  \nonumber\\
&&r=\ln\cot\psi_2+i(\varphi_{21}-\varphi_{22})\,,\nonumber\\
&&w=\ln\cos \psi_1-i\ln\cos \psi_2-\varphi_{21}-i\, \varphi_{11}\,,
\end{eqnarray}
and the potential is well defined as long as $ \psi_1,\,\psi_2\neq 0$.

It is not hard to relate the present expressions to those obtained in previous section
for $SU(2)\times U(1)\subset SU(2)
\times SU(2)$ by taking an appropriate limit of the above. {\em E.g.} taking $\psi_1 \rightarrow 0$ in eq.~(\ref{s3s3V}) results in the $SU(2)\times U(1)$ potential in eq.~(\ref{piano1}). Which is another way to see that $SU(2)\times SU(2)$ degenerates at the endpoints $\psi_1=0$ ($\psi_2=0$) or $\psi_1=\pi/2$ ($\psi_2=\pi/2$).  

To conclude this section about the complex geometry of $SU(2)\times SU(2)$, let us turn to the isometries, $g \rightarrow h_-gh_+\,$, where we take,
\begin{eqnarray}
h_\pm= \left(
\begin{array}{cc}
e^{i \varepsilon_\pm^1 \sigma_3}&0\\
0& e^{i \varepsilon_\pm^2 \sigma_3}
\end{array}\right).
\end{eqnarray}
Under this the coordinates transform as,
\begin{eqnarray}
l \rightarrow  l+ \phi_- + \bar{ \phi }_+,\qquad r \rightarrow r+ \phi_+-\bar \phi_+,
\qquad z \rightarrow  z+ i\bar \phi_--i\bar \phi_+,
\end{eqnarray}
where,
\begin{eqnarray}
\phi_\pm\equiv \varepsilon_\pm^1+i\, \varepsilon^2_\pm\,.
\end{eqnarray}
These transformations are compatible with the constraints provided,
\begin{eqnarray}
\ID_+ \phi_-=\bar \ID_+ \phi_-=\ID_- \phi_-=0,\qquad
\ID_- \phi_+=\bar \ID_- \phi_+=\ID_+ \phi_+=0,
\end{eqnarray}
which once more implies that $ \partial _\pp\, \phi_-= \partial _= \phi_+=0$. It is straightforward to check that the model is indeed invariant under this.

\subsection{Generalized K\"ahler geometry and type changing}\label{sect:52}
Next, we take a closer look at the generalized K\"ahler geometry of this model. In order to proceed in a structured way we will treat the semi-chiral + chiral and semi-chiral + twisted chiral parametrization in a different subsection. But the analysis for both parametrizations follows a completely analogous pattern:
\begin{itemize}
\item[(1)] First we compute the expressions for both pure spinors, given the generalized K\"ahler potentials in (\ref{s3s3V}) or (\ref{s3s3V2}) respectively.
\item[(2)] Then we compute the Mukai pairings $(\phi_+, \bar \phi_+)$ and $(\phi_-, \bar \phi_-)$ and observe that there are points in the manifold where one of the two Mukai pairings vanishes. This signals the occurrence of type changing loci.  
\item[(3)] Next, we introduce complex coordinates for each type changing locus seperately such that a type changing locus can be unambiguously expressed in terms of the complex coordinates (instead of the real coordinates). In some cases, these coordinates will diagonalize $J_+$.
\item[(4)] We simplify the expressions for the pure spinors by performing a $b$-transform\footnote{Given that the expressions are more involved in six dimensions, we will not bother the reader by writing down the expressions for $b$, $\Lambda_+$ and $\Lambda_-$ explicitly.}. The resulting two-forms in the exponent of the pure spinors $\phi_+$ and $\phi_-$ are then called $\Lambda_+$ and $\Lambda_-$ respectively. 
\item[(5)] The pure spinors are rescaled in such a way that they reveal type changing explicitly (i.e.~a symplectic-like pure spinor outside the locus and a complex-like pure spinor in the type changing point.)
\end{itemize}

\subsubsection{The semi-chiral and chiral parametrization}\label{sect:521}
Given the generalized K\"ahler potential (\ref{s3s3V}) for the semi-chiral and chiral parametrization, we find the following two pure spinors,
\begin{eqnarray}
\phi_+ &=& d \bar z \wedge e^{i\, \Omega^+ + \Xi^+},\\
\phi_- &=& e^{i\, \Omega^- + \Xi^-},
\end{eqnarray}
with $i\, \Omega^+ + \Xi^+$ given by,
\begin{eqnarray}
i\, \Omega^+ + \Xi^+ = \frac{k}{8 \pi} \left( -\frac{2}{1+ e^{i\, (l - \bar l ) + z + \bar z}} dl \wedge d \bar l - dl\wedge d\bar r -\frac{2 i}{1+ e^{-i\, (l - \bar l) - z - \bar z}} dl\wedge d \bar z\right. \nonumber \\  - d\bar l \wedge d r + \frac{2 i}{ 1+ e^{-i\,(l - \bar l) - z -\bar z}} d\bar l \wedge d z + \frac{4i}{1+e^{-i\,(l - \bar l) - z -\bar z}} d \bar l \wedge d \bar z \nonumber\\
\left. + \frac{2}{1+e^{r+\bar r}} dr\wedge d\bar r - \frac{2}{1+ e^{-i\,(l - \bar l) - z -\bar z}} dz\wedge d\bar z  \right),
\end{eqnarray}
and $i\, \Omega^- + \Xi^-$  given by,
\begin{eqnarray}
i\, \Omega^- + \Xi^- = \frac{k}{8 \pi} \left( \frac{2}{1+ e^{i\, (l-\bar l) + z + \bar z}} dl\wedge d\bar l - dl\wedge d\bar r  - \frac{2i}{1+ e^{-i(l-\bar l) - z - \bar z}} dl \wedge d\bar z\right. \nonumber \\ + 3 d\bar l \wedge d r - \frac{2i}{1+e^{-i(l-\bar l) - z - \bar z}} d\bar l \wedge d z \nonumber \\ \left. + \frac{2}{1+ e^{r+ \bar r}}  dr \wedge d \bar r  - \frac{2}{1+e^{-i\, (l-\bar l) - z - \bar z}} dz \wedge d \bar z \right).
\end{eqnarray}
With these expressions for the pure spinors we calculate the Mukai pairings,
\begin{eqnarray}
(\phi_+, \bar \phi_+) &=& \frac{k^2}{4 \pi^2} \frac{1}{(1+e^{r+\bar r}) (1+ e^{i\, l - i\, \bar l + z +\bar z})} =  \frac{k^2}{4 \pi^2} \cos^2 \psi_1 \, \cos^2 \psi_2 , \nonumber\\
(\phi_-, \bar \phi_-) &=& - \frac{k^3}{8 \pi^3} \frac{e^{i\, l - i\, \bar l + r+ \bar r + z + \bar z}}{(1+e^{r+\bar r}) (1+ e^{i\, l - i\, \bar l + z +\bar z})} =  - \frac{k^3}{8 \pi^3} \sin^2 \psi_1 \, \sin^2 \psi_2 .
\end{eqnarray}
We notice two important things from this calculation: the generalized Calabi-Yau condition (\ref{CYcd1}) is not satisfied in this patch, while the weaker version in eq.~(\ref{CYcd2}) is satisfied. Secondly, we find three loci where one of the Mukai pairings vanishes and thus where type changing occurs: $\psi_1 = \pi/2 = \psi_2$, $\psi_1 = 0$ and $\psi_2=0$. For each locus we must now introduce specific complex coordinates in which type changing becomes lucid and explicit. 

\vspace{0.1in}
\noindent \underline{\em i. Type changing locus 1} :  $\psi_1 = 0 = \psi_2$ \label{sect:5211}\\ 
We introduce the following complex coordinates,
\begin{eqnarray}
z^1 = e^l,\qquad z^2 = e^{-i\, \bar l + i\, r - \ln \big(1+ e^{i\, l - i\, \bar l + z + \bar z} \big)}, \qquad z^3 = e^{z + i\, l},
\end{eqnarray}
and complex conjugates, such that $J_+$ is diagonalized to $+i$ on $dz^1, dz^2$ and $dz^3$ and $-i$ on $d\bar z^1$, $d\bar z^2$ and $d\bar z^3$. In these coordinates the type changing locus is located at $z^1 = 0 = z^3$. When we take $z^1$ and $z^3$ to zero, the second complex structure $J_-$ diagonalizes as well and we find that $J_+ = J_-$. This implies that the type of the pure spinors jumps from $(1,0)$ to $(3,0)$ when $z^1 = 0$ and $z^3=0$.  Let us take a closer look at the pure spinors. We first rewrite the pure spinors in the new complex coordinates and we simplify the expressions for both pure spinors by performing a $b$-transform,
\begin{eqnarray}
\phi_+ &=& i \sqrt{z^1 \bar z^1 z^2  \bar z^2 z^3  \bar z^3} \left( \frac{d\bar z^3}{\bar z^3} + i\, \frac{d \bar z^1}{\bar z^1} \right) \wedge e^{-b\wedge}\, e^{\Lambda_+}, \nonumber \\
\phi_- &=& i \sqrt{z^1 \bar z^1 z^2 \bar z^2 z^3  \bar z^3}\, e^{-b\wedge}\, e^{\Lambda_-},
\end{eqnarray} 
where one can easily show that $d\Lambda_+ = d\Lambda_- = db = H$. Writing down the two pure spinors after the $b$-transform (i.e.~without the $b$-dependent part) leaves us with two pure spinors that are $H$-closed.
 
In order to unravel type changing at $z^1 = 0 = z^3$, we should rescale the first pure spinor:
\begin{eqnarray}
\phi_+ = i\, \frac{2 \pi}{k} \bar z^1 \bar z^2 \bar z^3 \left( \frac{d\bar z^3}{\bar z^3} + i\, \frac{d \bar z^1}{\bar z^1} \right) \wedge e^{\Lambda_+}, \label{p1l1psp}
\end{eqnarray}
which reduces to the standard complex form $d\bar z^1\wedge d \bar z^2 \wedge d\bar z^3$ in $z^1 = 0 = z^3$. $\phi_-$ on the other hand is rescaled to the form
\begin{eqnarray}
\phi_- = e^{\Lambda_-}, \label{p1l1psm}
\end{eqnarray}
such that it takes on the standard form of a symplectic pure spinor in $z^1 = 0 = z^3$. One can now show that $\phi_-$ is still $H$-closed, but that $\phi_+$ is no longer $H$-closed. Let us now also calculate the Mukai pairings for the rescaled pure spinors,
\begin{eqnarray}
(\phi_+, \bar \phi_+) &=& \frac{1}{1+ z^1 \bar z^1 (z^2)^{-i} (\bar z^2)^i }\, \frac{1}{1+ z^3 \bar z^3},\\
(\phi_-, \bar \phi_-) &=& -\frac{k^3}{8 \pi^3}\, \frac{1}{ (z^2)^i (\bar z^2)^{-i}+ z^1 \bar z^1 }\, \frac{1}{1+ z^3 \bar z^3} \,\frac{1}{z^2 \bar z^2}.
\end{eqnarray} 
We observe that both Mukai pairings behave regularly also when $z^1 = 0 = z^3$, but that they do not satisfy the generalized Calabi-Yau condition (\ref{CYcd1}).

\vspace{0.1in}
\noindent \underline{\em ii. Type changing locus 2} : $\psi_1 = \pi/2$\label{sect:5212}\\
To investigate type changing in the second locus, we introduce the following complex coordinates,
\begin{eqnarray}
w^1 = e^{-i\, l}, \qquad w^2 = e^{-\bar r - i\, \bar z}, \qquad z^1 = e^{i\, z},
\end{eqnarray}
and their complex conjugates. In these coordinates the type changing occurs at $w^1 = 0$. Neither of the complex structures is diagonalized, but we observe that both $J_+$ and $J_-$ reduce to diagonal complex structures for $w^1 \rightarrow 0$:
\begin{eqnarray}
J_+ = {\rm diag} (+i, -i, +i, -i, +i, -i), \qquad J_- = {\rm diag}(-i, +i, -i, +i, +i, -i),
\end{eqnarray}
which shows that there are two complex directions for which $J_+ = -J_-$ and one complex direction for which $J_+ = J_-$. This implies that we require two twisted chiral superfields and one chiral superfield to parametrize the model at $w^1 = 0$ (or $\psi_1 = \pi/2$). The type of the pure spinors then jumps from $(1, 0)$ to $(1,2)$ at $w^1 = 0$. 

Let us first write down the pure spinors in the new complex coordinates and point out that we can rewrite the pure spinors by performing a b-transform,
\begin{eqnarray}
\phi_+ &=& -\, \sqrt{w^1 \bar w^1 w^2 \bar w^2 z^1 \bar z^1}\, \frac{d\bar z^1}{\bar z^1} \wedge e^{-b\wedge}\, e^{\Lambda_+} ,\\
\phi_- &=&i\, \sqrt{w^1 \bar w^1 w^2 \bar w^2 z^1 \bar z^1}\, e^{-b\wedge}\, e^{\Lambda_-},
\end{eqnarray}
and one can show that $d \Lambda_+ = d\Lambda_- = db = H$. Pure spinors of the form $e^{\Lambda_\pm}$ are no longer closed, but $H$-closed.

Next, we rescale the pure spinors to make the type changing explicit. The first pure spinor takes the form
\begin{eqnarray}
\phi_+ = d\bar z^1 \wedge e^{\Lambda_+} ,
\end{eqnarray}
while the second pure spinor is rescaled to the form,
\begin{eqnarray}
\phi_- = -i\, \frac{2 \pi}{k} \bar w^1 \bar w^2 e^{\Lambda_-}.
\end{eqnarray}
When we take $w^1\rightarrow 0$, we note that $\phi_+$ remains unchanged with respect to its type and that $\phi_+$ contains the pure spinor as given in the expression (\ref{Lampsimctcp2}), reflecting the remaining $S^3\times S^1$ factor at $w^1=0$. One can see explicitly that the group manifold is pinched down to a $S^3\times S^1$ at $w^1=0$ by taking $\psi_1=\pi/2$ in the expressions for the metric (\ref{S3S3met}) and the torsion three-form (\ref{S3S3tor}). The twisted chiral superfield is in this case played by $w^2$, while the r\^ole of the chiral superfield is played by $z^1$. At $w^1 = 0$ the second pure spinor takes on a more complex-like form (instead of a symplectic-like),
\begin{eqnarray}
\phi_- \big|_{w^1=0} = d \bar w^1 \wedge d \bar w^2 \wedge e^{\Lambda_- |_{w^1=0}},
\end{eqnarray}
where by $\Lambda |_{w^1=0}$ we mean those pieces of $\bar w^1 \bar w^2 \Lambda_-$ that do not vanish at $w^1 = 0$. It turns out that $d \bar w^2 \wedge e^{\Lambda |_{w^1=0}}$ is exactly the pure spinor as presented in eq.~(\ref{Lammsimctcp2}), so that we recover the $S^3\times S^1$ factor (parameterized by chiral superfield $z^1$ and twisted chiral superfield $w^2$) also in the second pure spinor. Again, we calculate the Mukai pairings for the rescaled pure spinors,
\begin{eqnarray}
(\phi_+, \bar \phi_+) &=& \frac{k^2}{4 \pi^2}\, \frac{1}{w^1 \bar w^1+(z^1)^{-i} (\bar z^1)^{i} }\, \frac{1}{w^2 \bar w^2+ z^1 \bar z^1}\, \frac{1}{z^1 \bar z^1},\\
(\phi_-, \bar \phi_-) &=& -\frac{k}{2 \pi} \, \frac{1}{1+ w^1 \bar w^1 (z^1)^{i} (\bar z^1)^{-i} }\, \frac{1}{w^2 \bar w^2+ z^1 \bar z^1}.
\end{eqnarray} 
We notice again that neither of the Mukai pairings vanishes in $w^1 = 0$ and that the generalized Calabi-Yau condition (\ref{CYcd1}) is not satisfied.

\vspace{0.1in}
\noindent \underline{\em iii. Type changing locus 3} : $\psi_2 = \pi/2$\label{sect:5213}\\
For the third locus we introduce the complex coordinates,
\begin{eqnarray}
w^1 = e^{-i\, l}, \qquad w^2 = e^{\bar l - \bar r - i\, \ln\big( 1+ e^{i\, l -i\, \bar l + z + \bar z} \big)}, \qquad z^1 = e^z,
\end{eqnarray}
and their complex conjugates, for which the complex structure $J_+$ diagonalizes to its standard diagonalized form ($+i$ on the holomorphic coordinates and $-i$ on the anti-holomorphic coordinates). In these complex coordinates the type changing locus is located at $w^2 = 0$. Before discussing type changing in this point, we  rewrite the pure spinors in terms of the new complex coordinates and we simplify the pure spinors by performing a $b$-transform,
\begin{eqnarray}
\phi_+ &=& i\, \sqrt{w^1 \bar w^1 w^2 \bar w^2 z^1   \bar z^1}\, \frac{d\bar z^1}{\bar z^1} \wedge e^{-b \wedge}\, e^{\Lambda_+},\nonumber\\
\phi_- &=&  i\, \sqrt{w^1 \bar w^1 w^2  \bar w^2  z^1 \bar z^1}\, e^{-b \wedge}\, e^{\Lambda_-},
\end{eqnarray}
where one can show that $d\Lambda_+ = d\Lambda_- = db = H$. After the $b$-transform (i.e.~without the $b$-dependent part) we find two pure spinors that are $H$-closed.

In the point $w^2 = 0$ the second complex structure $J_-$ reduces to diag($-i, i, -i, i, i, -i$), indicating that there are two (complex) directions for which $J_+ = - J_-$ and one (complex) direction for which $J_+ = J_-$. This implies that we require two twisted chiral superfields and one chiral superfield in the type changing locus $w^2 = 0$. The type itself then jumps from $(1,0)$ to $(1,2)$. In order to make the type changing more explicit, we rescale both pure spinors, such that they take the following form:
\begin{eqnarray}
\phi_+ &=& d\bar z^1 \wedge e^{\Lambda_+}, \nonumber\\
\phi_- &=& -i\, \frac{2 \pi}{k} \bar w^1 \bar w^2 e^{\Lambda_-}.
\end{eqnarray}
When taking $w^2$ to zero $\phi_+$ contains the pure spinor (without the $b$-transform) as given in expression(\ref{Lampsimctcp2}), where now $w^1$ plays the r\^ole of the twisted chiral superfield and $z^1$ the r\^ole of the chiral superfield. In the type changing locus $w^2=0$ we thus retrieve a $S^3\times S^1$ factor parametrized by a chiral and a twisted chiral superfield. The fact that the manifold is pinched down to $S^3\times S^1$ can also be seen by taking $\psi_2=\pi/2$ in the expressions for the metric (\ref{S3S3met}) and the torsion three-form (\ref{S3S3tor}).  $\phi_-$ on the other hand reduces to the following form,
\begin{eqnarray}
\phi_-\big|_{w^2 = 0} = d\bar w^1 \wedge d\bar  w^2 \wedge e^{\Lambda_-|_{w^2=0} },
\end{eqnarray}
where $e^{\Lambda_-|_{w^2=0}}$ represents those pieces of $w^{\bar 1}  w^{\bar 2} e^{\Lambda_-}$ which do not vanish when we take $w^2 \rightarrow 0$. It is not difficult to see that $d\bar w^1\wedge e^{\Lambda_-|_{w^2=0}}$ corresponds to the pure spinor as given in eq.~(\ref{Lammsimctcp2}). Hence, $w^1$ and $z^1$ form the correct complex coordinates to parameterize the $S^3\times S^1$ factor at $w^2 = 0$. To conclude the discussion about type changing in this patch, we calculate the Mukai pairings for the rescaled pure spinors,
\begin{eqnarray}
(\phi_+, \bar \phi_+) &=& \frac{k^2}{4 \pi^2}\, \frac{1}{w^2 \bar w^2+(w^1)^{i} (\bar w^1)^{-i} }\, \frac{1}{w^1 \bar w^1+ z^1 \bar z^1}\, \frac{1}{z^1 \bar z^1},\\
(\phi_-, \bar \phi_-) &=& -\frac{k}{2 \pi} \, \frac{1}{1+(w^1)^{-i} (\bar w^1)^{i}  w^2 \bar w^2}\, \frac{1}{w^1 \bar w^1+ z^1 \bar z^1} .
\end{eqnarray} 
Here we find that the Mukai pairings behave regularly for the rescaled pure spinors, also in $w^2=0$. Once more, the generalized Calabi-Yau condition (\ref{CYcd1}) is not satisfied.

 %%%%%%%%%%%%%%%%%%%%%%%%%%%%%%%%%%%%%%%%%%%%%%%%%%%%%%%%
\subsubsection{The semi-chiral and twisted chiral parametrization}\label{sect:522}
Starting from  the generalized K\"ahler potential (\ref{s3s3V2}) for the semi-chiral and twisted chiral parametrization, we obtain the following expressions for the pure spinors,
\begin{eqnarray}
\phi_+ &=& e^{i\, \Omega^+ + \Xi^+},\\
\phi_- &=& d\bar w \wedge e^{i\, \Omega^- + \Xi^-},
\end{eqnarray}
with, 
\begin{eqnarray}
i\, \Omega^+ + \Xi^+ = \frac{k}{8 \pi} \left( \frac{2}{1+ e^{i\,(l - \bar l)+w + \bar w}} dl\wedge d\bar l - dl \wedge dr + \frac{i}{1+e^{-i\,(l-\bar l) - w - \bar w}} dl \wedge dw   \right. \nonumber \\  - \frac{i}{1+e^{-i\,(l-\bar l) - w - \bar w}} dl \wedge d\bar w + 3 d\bar l \wedge d\bar r - \frac{3i}{1+e^{-i\,(l-\bar l) - w - \bar w}} d\bar l \wedge d w \nonumber \\  - \frac{i}{1+ e^{-i\,(l-\bar l) - w - \bar w}} d\bar l \wedge d \bar w - \frac{2}{1+e^{r + \bar r}} dr \wedge d\bar r \nonumber \\ \left. - \frac{2i}{1+ e^{-i\,(l-\bar l ) - w - \bar w}} dw \wedge d \bar w  \right),
\end{eqnarray}
and, 
\begin{eqnarray}
i\, \Omega^- + \Xi^- = \frac{k}{8 \pi} \left( -\frac{2}{1+e^{i\,(l-\bar l) + w + \bar w }} dl \wedge d\bar l - dl\wedge dr + \frac{i}{1+e^{-i\,(l-\bar l) - w - \bar w}} dl\wedge dw \right. \nonumber\\  + \frac{3}{1+ e^{-i\,(l-\bar l) - w - \bar w}} d l \wedge d\bar w - d\bar l \wedge d \bar r + \frac{i}{1+e^{-i\,(l-\bar l) - w - \bar w}} d\bar l \wedge dw \nonumber  \\   - \frac{i}{1+e^{-i\,(l- \bar l ) - w - \bar w}} d\bar l \wedge d \bar w - \frac{2}{1+e^{r + \bar r}} d r\wedge d\bar r \nonumber \\ \left. + \frac{2}{1+e^{-i\,(l-\bar l ) - w -\bar w}} dw \wedge d \bar w  \right) .
\end{eqnarray}
Let us also calculate the Mukai pairings for these pure spinors:
\begin{eqnarray}
(\phi_+, \bar \phi_+) &=& \frac{k^3}{8 \pi^3} \frac{e^{i\, l - i \, \bar l + r + \bar r+ w + \bar w}}{(1+e^{r+\bar r})(1+e^{i\, l -i\, \bar l + w + \bar w})} = \frac{k^3}{8 \pi^3} \cos^2 \psi_1 \, \cos^2\psi_2,\\
(\phi_-, \bar \phi_-) &=& -\frac{k^2}{4 \pi^2} \frac{1}{(1+e^{r+\bar r})(1+e^{i\, l -i\, \bar l + w + \bar w})} = - \frac{k^2}{4 \pi^2} \sin^2 \psi_1 \, \sin^2 \psi_2  .
\end{eqnarray}
Also in this patch we observe two important points: the generalized Calabi-Yau condition (\ref{CYcd1}) is once more not satisfied, while the weaker version (\ref{CYcd2}) is. And we find three type changing loci where one of the Mukai pairings vanishes: $\psi_1 = \pi/2 = \psi_2$, $\psi_1 = 0$ and $\psi_2 = 0$. Hence, we must find again complex coordinates for each locus (separately) such that the type changing of the pure spinors becomes manifest.

\vspace{0.1in}
\noindent \underline{\em i. Type changing locus 1} :  $\psi_1 = \pi/2 = \psi_2$ \label{sect:5221}\\ 
We first introduce the following complex coordinates,
\begin{eqnarray}
w^1 = e^l, \qquad w^2 = e^{-i\, \bar l + i\, r - \ln\big(1+ e^{i\, l - i\, \bar l + w +\bar w} \big)}, \qquad w^3 = e^{w+i\, l},
\end{eqnarray}
and their complex conjugates, such that $J_+$ is a diagonalized complex structure ($+i$ on $dw^1$, $dw^2$ and $dw^3$, and $-i$ on $d\bar w^1$, $d\bar w^2$ and $d\bar w^3$). In these complex coordinates the type changing locus is situated at $w^1 = 0 = w^3$, and in that particular point also $J_-$ diagonalizes such that $J_- = - J_+$. This last relation implies that the type of the pure spinors jumps from $(0,1)$ to $(0,3)$. So let us have a look at the pure spinors themselves. First, we have to take into account that they transform as densities when we perform a coordinate transformation,
\begin{eqnarray}
\phi_+ &=& \sqrt{w^1 \bar w^1 w^2 \bar w^2 w^3 \bar w^3 } \, e^{-b\wedge}\, e^{\Lambda_+},\\
\phi_- &=& \sqrt{w^1 \bar w^1 w^2 \bar w^2 w^3 \bar w^3 } \, \left( \frac{d \bar w^3}{\bar w^3} + i\, \frac{d \bar w^1}{\bar w^1} \right)\wedge e^{-b\wedge}\, e^{\Lambda_-}
\end{eqnarray}
At the same time we simplify the expressions for the pure spinors using a $b$-transform, and one can show that $d\Lambda_+ = d \Lambda_- = db = H$. Hence, after the $b$-transform the pure spinors are $H$-closed.

In order to make the type changing more explicit, we rescale the pure spinors such that they take the following form
\begin{eqnarray}
\phi_+ &=& e^{\Lambda_+}, \\
\phi_- &=& i\, \frac{2 \pi}{k} \bar w^1 \bar w^2 \bar w^3 \left( \frac{d \bar w^3}{\bar w^3} + i\, \frac{d \bar w^1}{\bar w^1} \right) \wedge  e^{\Lambda_-}.
\end{eqnarray}
When we take $w^1$ and $w^3$ to zero, we observe that $\phi_+$ remains a symplectic-type pure spinor, while $\phi_-$ reduces to the standard complex-like pure spinor $d\bar w^1 \wedge d\bar w^2 \wedge d\bar w^3$. $\phi_+$ is still $H$-closed, but $\phi_-$ is no longer $H$-closed. We also calculate the Mukai pairings for the rescaled pure spinors,
\begin{eqnarray}
(\phi_+, \bar \phi_+) &=& - \frac{k^3}{8 \pi^3}\, \frac{1}{(w^2)^i (\bar w^2)^{-i} + w^1 \bar w^1} \, \frac{1}{1+ w^3 \bar w^3}\, \frac{1}{w^2 \bar w^2}, \\
(\phi_-, \bar \phi_-) &=& \frac{1}{1+ w^1 \bar w^1 (w^2)^{-i} (\bar w^2)^{i}}\, \frac{1}{1+w^3 \bar w^3},
\end{eqnarray}
which never vanish in these coordinates, not even in $w^1 = 0 = w^3$. They also confirm once again that in this patch for $SU(2)\times SU(2)$ the generalized Calabi-Yau condition (\ref{CYcd1}) is not satisfied.

\vspace{0.1in}
\noindent \underline{\em ii. Type changing locus 2} : $\psi_1 = 0$\label{sect:5222}\\
To clarify the type changing in $\psi_1 = 0$ we introduce the complex coordinates,
\begin{eqnarray}
z^1 = e^{-i\, l}, \qquad z^2 = e^{-r - i\, \bar w}, \qquad w^1 = e^{i\, w},
\end{eqnarray}
and their complex conjugates. In these coordinates neither of the complex structures is diagonalized, but when we consider the type changing locus ($w^1 = 0$), we observe that both complex structures are diagonalized,
\begin{eqnarray}
J_+ = {\rm diag}(+i, -i, +i, -i, +i, -i), \qquad J_- = {\rm diag}(+i, -i, +i, -i, -i, +i).
\end{eqnarray}
We observe that there are two (complex) directions along which the complex structures are equal to each other, and one (complex) directions along which they are opposite. This indicates that we require two chiral superfields and one twisted chiral superfield to parameterize the manifold at $z^1 = 0$. And it also implies that the type of the pure spinors jumps from $(0,1)$ to $(2,1)$. Looking explicitly at the pure spinors,
\begin{eqnarray}
\phi_+ &=& \sqrt{w^1 \bar w^1 z^1 \bar z^1 z^2 \bar z^2} \wedge e^{-b\wedge}\, e^{\Lambda_+}\, \\
\phi_- &=& i\, \sqrt{w^1 \bar w^1 z^1 \bar z^1 z^2 \bar z^2}\, \frac{d\bar w^1}{\bar w^1} \wedge e^{-b \wedge}\, e^{\Lambda_-},
\end{eqnarray}
where one can easily show that $d\Lambda_+ = d \Lambda_- = db = H$, such that the pure spinors (after the $b$-transform) are $H$-closed (instead of closed).

Type changing becomes manifest by rescaling the pure spinors to the following form:
\begin{eqnarray}
\phi_+ &=& -i\, \frac{2 \pi}{k} \bar z^1 \bar z^2\, e^{\Lambda_+}, \nonumber \\ 
\phi_- &=& d\bar w^1 \wedge e^{\Lambda_-}.
\end{eqnarray}
Taking $z^1$ to zero, we find that $\phi_-$ remains of the same type and contains the pure spinor as given in the expression (\ref{Lammsimctcp1}). Hence, we find that the manifold reduces to a $S^3\times S^1$ parametrized by $z^2$ (chiral superfield) and $w^1$ (twisted chiral superfield). This is consistent with the metric and torsion 3-form one obtains when filling in $\psi_1 = 0$ in (\ref{S3S3met}) and (\ref{S3S3tor}) respectively. The other pure spinor $\phi_+$ does change type and becomes a more complex-like pure spinor,
\begin{eqnarray}
\phi_+ \big|_{z^1 = 0} = d \bar z^1 \wedge d \bar z^2 \wedge e^{\Lambda_- |_{z^1=0}},
\end{eqnarray} 
where $d \bar z^2 \wedge e^{\Lambda_- |_{z^1=0}}$ exactly corresponds to the pure spinor as given in eq.~(\ref{Lampsimctcp1}) in terms of $z^2$ (as chiral superfield) and $w^1$ (as twisted chiral superfield). One can show that $\phi_-$ is still $H$-closed, while $\phi_+$ is no longer $H$-closed. Next, we calculate the Mukai pairings for the rescaled pure spinors,
\begin{eqnarray}
(\phi_+, \bar \phi_+) &=& - \frac{k}{2 \pi} \frac{1}{1+ z^1 \bar z^1 (w^1)^i (\bar w^1)^{-i}}\, \frac{1}{z^2 \bar z^2 + w^1 \bar w^1} ,\\
(\phi_-, \bar \phi_-) &=&\frac{k^2}{4 \pi^2} \frac{1}{ (w^1)^{-i} (\bar w^1)^{i}+ z^1 \bar z^1}\, \frac{1}{z^2 \bar z^2 + w^1 \bar w^1}\, \frac{1}{w^1 \bar w^1}.
\end{eqnarray}
In these complex coordinates, we can see clearly again that the Mukai pairings do not vanish, not even in $z^1 = 0$ and that the generalized Calabi-Yau condition (\ref{CYcd1}) is not satisfied.

\vspace{0.1in}
\noindent \underline{\em iii. Type changing locus 3} : $\psi_2 = 0$\label{sect:5223}\\
Finally, we introduce the following complex coordinates,
\begin{eqnarray}
z^1 = e^{-i\, l}, \qquad z^2 = e^{\bar l - r - i\, \ln \big( 1 + e^{i\, l - i\, \bar l + w + \bar w } \big)}, \qquad w^1 = e^w,
\end{eqnarray}
and their complex conjugates to describe the third type changing locus in this patch. In these coordinates $J_+$ is diagonalized to the standard form of a complex structure ($+i$ on $dz^1$, $dz^2$, $dw^1$ and $-i$ on $d\bar z^1$, $d\bar z^2$ and $d\bar w^1$). The type changing locus is situated at $z^2 = 0$, in which also the other complex structure $J_-$ diagonalizes to the form diag($+i, -i, +i, -i, -i, +i)$. Hence, in $z^2=0$ there are two (complex) directions for which $J+ = J_-$ and one (complex) direction for which $J_+ = -J_-$. This implies that we require two chiral superfields and one twisted chiral superfield and that the type of the pure spinors jumps from $(0,1)$ to $(2,1)$. Let us thus consider the pure spinors explicitly,
\begin{eqnarray}
\phi_+ &=& \sqrt{z^1 \bar z^1 z^2 \bar z^2 w^1 \bar w^1}\, e^{-b\wedge} \, e^{\Lambda_+} ,\\
\phi_- &=& \sqrt{z^1 \bar z^1 z^2 \bar z^2 w^1 \bar w^1}\, \frac{d \bar w^1}{\bar w^1} \wedge e^{-b \wedge} \, e^{\Lambda_-}, 
\end{eqnarray}
where one can show that $d\Lambda_+ = d \Lambda_- = db = H$. Leaving out the $b$-transform this implies that the two pure spinors are $H$-closed. Type changing really becomes manifest after rescaling these pure spinors,
\begin{eqnarray}
\phi_+ &=& - i\, \frac{2 \pi}{k} \bar z^1 \bar z^2  e^{\Lambda_+} ,\\
\phi_- &=& d \bar w^1 \wedge e^{\Lambda_-}.
\end{eqnarray} 
If we then look at what happens in $z^2=0$, we observe that $\phi_+$ becomes more complex-like,
\begin{eqnarray}
\phi_+\big|_{z^2=0} &=& d \bar z^1 \wedge d\bar z^2\wedge e^{\Lambda_+ |_{z^2=0} }, 
\end{eqnarray}
such that $d\bar z^1 \wedge e^{\Lambda_+ |_{z^2=0}}$ corresponds to the pure spinor given in eq.~(\ref{Lampsimctcp1}), where $z^1$ plays the r\^ole of the chiral superfield and $w^1$ the r\^ole of the twisted chiral superfield. The second pure spinor behaves regularly, remains of the same type and contains the pure spinor as given in expression (\ref{Lammsimctcp1}). Hence, we can conclude that also in $z^2=0$ the manifold looks like a $S^3\times S^1$ factor, now parametrized by $z^1$ and $w^1$. We also note that $\phi_+$ is no longer $H$-closed, while $\phi_-$ remains $H$-closed. To conclude this discussion, we calculate the Mukai pairings for the rescaled pure spinors:
\begin{eqnarray}
(\phi_+, \bar \phi_+) &=& - \frac{k}{2 \pi} \frac{1}{1+ (z^1)^{-i} (\bar z^1)^{i} z^2 \bar z^2 }\, \frac{1}{z^1 \bar z^1 + w^1 \bar w^1} ,\\
(\phi_-, \bar \phi_-) &=&\frac{k^2}{4 \pi^2} \frac{1}{ (z^1)^{i} (\bar z^1)^{-i}+ z^2 \bar z^2}\, \frac{1}{z^1 \bar z^1 + w^1 \bar w^1}\, \frac{1}{w^1 \bar w^1}.
\end{eqnarray} 
Both Mukai pairings behave regularly and do not vanish in $z^2=0$ and they show once more that the generalized Calabi-Yau condition (\ref{CYcd1}) is not satisfied for $SU(2)\times SU(2)$.

\section{Discussion and conclusions}
From the present paper it is clear that even-dimensional reductive group manifolds provide a very 
explicit and manageable class of models where various aspects of generalized K\"ahler geometry 
can be studied. Only $SU(2)\times U(1)$ allows for commuting complex structures and thus can 
be described solely in terms of chiral and twisted chiral superfields. Generically semi-chiral fields 
will be omnipresent. The current examples show that the presence of semi-chiral superfields
seems to be closely connected with the occurence of type changing. In fact we expect that type
changing is a generic feature for reductive group manifolds, a fact already recognized in 
\cite{Gualtieri:2003dx}. 

This can easily be 
demonstrated when we focus on the subclass of models for which the supersymmetry is 
enhanced to 
$N=(4,4)$. These models were identified and classified in \cite{Spindel:1988nh} and are well known 
non-trivial examples of hypercomplex manifolds. The result is 
that $N=(4,4)$ is possible on $U(1)^4$, $SU(2)\times U(1)$ and on groups which can be written 
as the product of factors  $W\times SU(2)\times U(1)$ where $W$ is a Wolf space (a symmetric 
quaternionic space). 
A full list of the hypercomplex group manifolds is 
$SU(2n+1)$, $SU(2n)\times U(1)$, $SO(4n)\times \big(U(1)\big)^{2n}$, 
 $SO(4n+2)\times \big(U(1)\big)^{2n-1}$,  $SO(2n+1)\times \big(U(1)\big)^{n}$,
 $Sp(2n)\times \big(U(1)\big)^{n}$, $G_2\times \big(U(1)\big)^2$,    
$F_4\times \big(U(1)\big)^4$, $E_6\times \big(U(1)\big)^2$, $E_7\times \big(U(1)\big)^7$ 
and $E_8\times \big(U(1)\big)^8$. 
A group manifold of this type has a two-sphere worth 
of complex structures on the Lie 
algebra. Choosing two non-coinciding, non-antipodal points 
for constructing the right and left invariant complex on the group guarantees that 
$\ker [J_+,J_-]=0$ and is therefore necessarily fully described in terms of semi-chiral 
multiplets. From eqs.~(\ref{globa2}) and (\ref{globa1}) one would then naively deduce that the 
Kalb-Ramond form is globally well defined and as a consequence that the canonical 3-form on the 
group is not only closed but exact as well, which is obviously impossible. The only way out is 
to assume the existence of 
loci where the type changes -- {\em i.e.} where $\ker [J_+,J_-]\neq 0$ -- 
implying that the two-form $ \Omega$, eq.~(\ref{globa1}),  is not well defined anymore. This was 
explicitly illustrated for $SU(2)\times U(1)$.  

We also point out that type changing occurs for both parametrizations of $SU(2)\times SU(2)$, indicating 
that type changing does not exclusively occur in models which possess an enhanced $N= (4,4)$ 
supersymmetry or in models which are purely parametrized by semi-chiral superfields. In the first patch 
(the semi-chiral + chiral parametrization) we found three different loci where the type changes, which in 
this case implies that the two-form $\Omega^-$ in eq.~(\ref{omglob}) is not well defined in these three 
points. Also for the second patch (the semi-chiral + twisted chiral parametrization) the three different type 
changing loci indicate that there is a two-form which is not well defined in every point of the manifold, 
namely the two-form $\Omega^+$ in eq.~(\ref{opglob}). An overview of the type changing can be found 
in table \ref{capTab1} for $SU(2)\times U(1)$ and in table \ref{capTab2} for $SU(2)\times SU(2)$. For 
models parametrized by all three types of superfield we know that there does not exist a plausible globally 
defined two-form. Hence, we might guess that type changing in these models is closely linked to this 
inexistence of a globally defined two-form. 

\begin{table}
\begin{center}
\begin{tabular}{c@{\qquad}c}
\hline $S^3\times S^1$ & locus \\
\hline \hline \\
patch 1 (\ref{Vsemilog}) & $\psi = \pi/2$: $(0,0) \rightarrow (0,2)$
\\
&\\
\hline \hline \\
patch 2 (\ref{Vsemilog2}) & $\psi = 0 $: $(0,0) \rightarrow (2,0)$\\
& \\
\hline
\end{tabular}\begin{picture}(0,0) \put(-235,-30){\begin{tikzpicture}  \draw[<->] (0,0) arc (270:90:0.8cm);\end{tikzpicture}} \put(-280, -5){mirror} \put(-290,-15){symmetry}  \end{picture}
\caption{Summary of type changing for $SU(2)\times U(1)$. \label{capTab1}}
\end{center}
\end{table}
\begin{table}
\begin{center}
\begin{tabular}{c@{\qquad}c@{\qquad}c@{\qquad}c}
\hline $S^3\times S^3$ & locus 1 & locus 2 & locus 3 \\
\hline \hline \\
patch 1 (\ref{s3s3V}) & $\psi_1 = 0 = \psi_2 $:  & $\psi_1= \pi/2$:  & $\psi_2= \pi/2$: \\
&$ (1,0)\rightarrow (3,0)$ & $(1,0)\rightarrow (1,2)$& $(1,0)\rightarrow (1,2)$\\
& & &\\
\hline \hline \\
patch 2 (\ref{s3s3V2}) & $\psi_1 = \pi/2 = \psi_2 $: & $\psi_1 = 0$:    &  $\psi_2= 0$:  \\
&  $(0,1)\rightarrow (0,3)$  & $(0,1)\rightarrow (2,1)$& $(0,1)\rightarrow (2,1)$\\
& & & \\
\hline
\end{tabular}\begin{picture}(0,0) \put(-375,-32){\begin{tikzpicture}  \draw[<->] (0,0) arc (270:90:1.1cm);\end{tikzpicture}} \put(-391, 0){mirror} \put(-400,-10){symmetry}  \end{picture}
\caption{Summary of type changing for $SU(2)\times SU(2)$ \label{capTab2}}
\end{center}
\end{table}

The generalized Calabi-Yau conditions on the 
generalized K\"ahler potential in $N=(2,2)$ superspace language 
were studied in \cite{Hull:2010sn}. The conditions found there, and 
rederived in eq.~(\ref{CYcd1}) are (apparently only mildly) stronger than the conditions for 1-loop UV 
finiteness calculated in 
\cite{Grisaru:1997pg}, and given in eq.~(\ref{CYcd1}).  As argued in \cite{Hull:2010sn}, this might
be a sign that the vanishing of the $\beta$-function is necessary but not sufficient to guarantee full 
(super)conformal invariance at the quantum level (see \cite{Hull:1985rc}). All of the WZW-models on 
reductive even-dimensional group manifolds are actually $N=(2,2)$ superconformally invariant at the 
quantum level. However they do not necessarily 
provide consistent supergravity backgrounds which gets reflected in 
the generalized Calabi-Yau conditions which are not satisfied. While the pure spinors for the $SU(2)\times 
U(1)$ background in terms of a chiral and a twisted chiral superfield do satisfy the generalized Calabi-Yau 
condition, a more careful investigation shows that the pure spinors are not globally defined in this case.

In the present paper we presented several examples -- $SU(2)\times U(1)$ in terms of a single 
semi-chiral multiplet and $SU(2)\times SU(2)$ in terms of a semi-chiral and one twisted chiral or chiral 
superfield -- where the closed pure spinors were constructed  and where the UV-finiteness condition 
eq.~(\ref{CYcd2}) is satisfied. All these examples correspond
to cases where type changing loci exist where the pure spinors were ill defined in the type changing loci. 
A manifestation of this is that some of the Mukai
pairings are degenerate in the type changing loci. Upon the introduction of appropriate complex 
coordinates
we could rescale the pure spinors so that they behave well everywhere even when the type changes. The
resulting pure spinors are however not closed anymore. These examples indicate that further study of
generalized K\"ahler geometry in the 
presence of type-changing loci  is called for. The explicit examples developed in the current paper might serve as a guide line in the search
for a definition which encompasses the problems signaled here. Finding a covariant -- in the context of
generalized K\"ahler geometry -- definition of the UV-finiteness condition in eq.~(\ref{CYcd2})
could certainly be an interesting first step. Furthermore, it is clear that in order to make progress in
connecting $N=(2,2)$ supersymmetric $\sigma$-model results to solutions of the supergravity equations 
of motion a thorough study of the dilaton coupling in $N=(2,2)$ superspace -- including the $\beta$-
functions -- is unavoidable. Work in this direction is currently in progress. 

Another interesting feature which appeared here is related to the global formulation of these 
models. Whenever the $\sigma$-model can be described in terms of chiral and twisted chiral coordinates 
only -- {\em i.e.} whenever $J_+$ and $J_-$ commute -- then the generalized K\"ahler potentials on the 
overlap of two coordinate patches are related by a generalized K\"ahler transformation (see {\em e.g.}
the discussion in \cite{Hull:2008vw}). Once semi-chiral 
coordinates are present as well -- {\em i.e.} when $\mbox{im}[J_+,J_-]g^{-1}\neq 0$ -- the situation 
becomes 
more involved and on the overlap of two coordinate patches an additional Legendre transform might be 
required in order to relate the potentials. A clear example of this was provided in section 4.4.1 where it 
was shown that the potentials for $S^3\times S^1$ on the two patches $\psi\neq 0$ and $\psi\neq \pi/2$ 
are each others mirror transform. On the overlap of the two patches the potentials are related by a 
Legendre transform. With this in mind it would certainly be worthwhile to further develop the ideas 
introduced in \cite{Hull:2008vw}.

In view of their relative simplicity, one would expect that a systematic description of the local generalized 
K\"ahler geometry of even dimensional reductive Lie group manifolds should be feasible. While we leave 
this to further investigation a few comments are still 
in order here. Indeed we saw that a complex structure on a group is almost completely determined once a 
Cartan decomposition for the Lie algebra is chosen. The remaining freedom lies in the choice of the 
complex structure on the Cartan subalgebra. The arbitrariness in the choice of a Cartan decomposition  
is translated in the precise relation between the complex coordinates and the original coordinates
parameterizing the group. The superfield content is fully determined by the choice of $\IJ_-$ with respect
to $\IJ_+$. {\em E.g.} when choosing $\IJ_-=\IJ_+$ we expect the number of semi-chiral superfields to 
be maximal. 
Another natural choice is given by taking $\IJ_+$ and $\IJ_-$ to be equal on the roots and taking them 
with an opposite sign on the Cartan subalgebra. This choice probably minimizes the number of semi-chiral
fields needed. 

From the previous discussion it is clear that given an even-dimensional reductive group manifold, 
we can construct several generalized K\"ahler geometries of different types. This is a consequence of the 
freedom one has for choosing the second complex structure on the Lie algebra. {\em E.g.} for the 
simplest case -- $SU(2)\times U(1)$ -- we found two descriptions, one in terms of a chiral and twisted 
chiral superfield and another one in terms of a semi-chiral multiplet. As we showed in section  4.3 both 
descriptions are T-dual to each other.  One might suspect that
alternate formulations of the same model are linked to each other through a chain of T-dualities,
{\em i.e.} different generalized K\"ahler geometries for a given manifold might all be related 
through T-duality. However this cannot be true. In order to see this we turn once more to hyper-K\"ahler 
manifolds which allow for two generalized K\"ahler geometries: the usual K\"ahler one where 
one generalized complex structure is of the complex type while the other is symplectic and the one where 
both generalized complex structures are symplectic (see eq.~(\ref{gualt})). In order that both descriptions
are T-dual to each other one would expect an isometry to exist. However, K3 provides an example of a 
hyper-K\"ahler manifolds without any isometries so we do expect that here the two generalized K\"ahler 
geometries, one of type $(2,0)$ and the other of type $(0,0)$ are not T-dual to each other.
%Finally, thanks to their simplicity, WZW-models on even-dimensional reductive group manifolds provide a 
%perfect playground to study various
%aspects of D-branes in a generalized K\"ahler geometry. In particular we intend to study the (classical) 
%isotropic and co-isotropic D-brane configurations constructed in \cite{Sevrin:2009na} at the quantum 
%level, read: in a generalized Calibi-Yau setting.

\acknowledgments

It is a pleasure to thank Nigel Hitchin, Tor Lowen, Jan Troost and especially Martin Ro\v cek for useful 
discussions and suggestions. AS thanks the Simons Center for Physics and Geometry for hospitality 
while part of this work was done.
AS and DT are supported in part by the Belgian Federal Science Policy Office
through the Interuniversity Attraction Pole P6/11, and in part by the
``FWO-Vlaanderen'' through the project G.0114.10N. WS was supported in part by the FWO through an 
FWO-aspirant fellowship and by the "Research Centre Elementary Forces and Mathematical Foundations" 
(EMG) at the Johannes Gutenberg-Universit\"at Mainz.

\appendix

\section{Conventions, notations and identities}\label{app conv}
The conventions used in the present paper are essentially the same as
those in \cite{Sevrin:2009na}.

We denote the worldsheet coordinates by $ \tau,\sigma \in\IR$, and the worldsheet light-cone coordinates are
defined by,
\begin{eqnarray}
\sigma ^\pp= \tau + \sigma ,\qquad \sigma ^== \tau - \sigma .\label{App1}
\end{eqnarray}
The $N=(1,1)$ (real) fermionic coordinates are denoted by $ \theta ^+$ and $ \theta ^-$ and the
corresponding derivatives satisfy,
\begin{eqnarray}
D_+^2= - \frac{i}{2}\, \partial _\pp \,,\qquad D_-^2=- \frac{i}{2}\, \partial _= \,,
\qquad \{D_+,D_-\}=0.\label{App2}
\end{eqnarray}
The $N=(1,1)$ integration measure is explicitely given by,
\begin{eqnarray}
\int d^ 2 \sigma \,d^2 \theta =\int d\tau \,d \sigma \,D_+D_-.
\end{eqnarray}
Passing from $N=(1,1)$ to $ N=(2,2)$ superspace requires
the introduction of two more real fermionic coordinates $ \hat \theta ^+$ and $ \hat \theta ^-$
where the corresponding fermionic derivatives satisfy,
\begin{eqnarray}
\hat D_+^2= - \frac{i}{2} \,\partial _\pp \,,\qquad \hat D_-^2=- \frac{i}{2} \,\partial _= \,,
\end{eqnarray}
and again all other -- except for (\ref{App2}) -- (anti-)commutators do vanish.
The $N=(2,2)$ integration measure is,
\begin{eqnarray}
\int d^2 \sigma \,d^2 \theta \, d^2 \hat \theta =
\int d \tau\, d \sigma \,D_+D_-\, \hat D_+ \hat D_-.
\end{eqnarray}
Regularly a complex basis is used,
\begin{eqnarray}
\ID_\pm\equiv \hat D_\pm+i\, D_\pm,\qquad
\bar \ID_\pm\equiv\hat D_\pm-i\,D_\pm,
\end{eqnarray}
which satisfy,
\begin{eqnarray}
\{\ID_+,\bar \ID_+\}= -2i\, \partial _\pp\,,\qquad
\{\ID_-,\bar \ID_-\}= -2i\, \partial _=,
\end{eqnarray}
and all other anti-commutators do vanish.

We denote the generators of a Lie algebra by $T_A$, $A\in\{1,\cdots,d\}$
and they satisfy the algebra $[T_A,T_B]=i\,f_{AB}{}^C\,T_C$. The
Cartan-Killing metric on the algebra, $\eta_{AB}$, is defined by,
\begin{eqnarray}
 \eta_{AB}\equiv -\frac{1}{\tilde h}\,f_{AC}{}^Df_{BD}{}^C,
\end{eqnarray}
with $\tilde h$ the dual Coxeter number of the Lie algebra. In general we
have,
\begin{eqnarray}
 \mbox{Tr}\big(T_AT_B\big)=x\,\eta_{AB},
\end{eqnarray}
with $x$ the index of the representation. One has that
$f_{ABC}=\eta_{CD}\,f_{AB}{}^D$ is fully anti-symmetric in its indices.

Denoting a group element by $g$ and a set of coordinates on the group by
$x^a$, $a\in\{1,\cdots ,d\}$, we introduce the left- and right-invariant
vielbeins $L_a^B$ and $R_a^B$,
\begin{eqnarray}
 g^{-1}dg=i\,L_a^B\,T_B\,dx^a,\qquad
 dgg^{-1}=i\,R_a^B\,T_B\,dx^a\,.
\end{eqnarray}
The  NSNS 3-form $H$ on the group is then given by,
\begin{eqnarray}
 H=\frac{k}{24\pi x}\,\mbox{Tr}\,dgg^{-1}\wedge dgg^{-1}\wedge dgg^{-1}=
 \frac{k}{48\pi }R_a^DR_b^ER_c^Ff_{DEF}\,dx^a\wedge dx^b\wedge dx^c,
\end{eqnarray}
where $k\in\IN$. The metric on the group is,
\begin{eqnarray}
 g_{ab}=-\frac{k}{8\pi  x}\,\mbox{Tr}\,\partial _a g g^{-1}\partial _b gg^{-1}=
 \frac{k}{8\pi }R_a^C\,R_b^D\,\eta_{CD}\,.
\end{eqnarray}

%: refs

\end{document}